\documentclass[floatfix,showpacs,amsmath,amssymb,letterpaper,groupaddresses,superscriptaddress]{article}
\usepackage{graphicx}
\usepackage{latexsym}
\usepackage{amsmath}
\usepackage{amssymb}
\usepackage{epsfig}
\usepackage{graphicx}
\usepackage{graphicx}
\usepackage{amssymb,amsmath,times}
\usepackage{hyperref}
\usepackage{color}

\def\be{\begin{equation}}
\def\ee{\end{equation}}
\def\ba{\begin{eqnarray}}
\def\ea{\end{eqnarray}}
\def\la{\langle}
\def\ra{\rangle}

\usepackage{epsfig}
\usepackage{graphicx}
\usepackage{multirow}
\usepackage{caption}
\begin{document}
\begin{center}
{\Large \bf General scheme for preparation of different topological states on the cluster states}\\
\vspace{1cm}Mohammad Hossein Zarei\footnote{email:mzarei92@shirazu.ac.ir}\\

\vspace{1cm}Physics Department, College of Sciences, Shiraz University, Shiraz 71454, Iran\\

\end{center}
\vskip 3cm
\begin{abstract}
Although it is well-known that all quantum states can be produced by single-qubit measurements on the cluster states, it is not a simple task to explicitly find which measurement patterns on the cluster states can generate different quantum states. In this paper, we introduce a general scheme to find measurement patterns corresponding to CSS topological states containing Kitaev's toric code states and color code states on different lattices and in different dimensions. Furthermore, we find a measurement pattern for generating non-abelian anyons where measurement-induced defects on a toric code state play the role of Ising anyons. We also support our scheme by a graphical notation where, by following a few simple graphical transformations, one will be able to convert a CSS topological state to a cluster state. Our scheme can also be used for experimental realization of anyons on cluster states.
\end{abstract}
\section{Introduction}
Topological order is a new concept which has originally emerged in condensed matter physics over the past decades in quantum Hall effect and quantum spin fluids \cite{hall, spin1, spin2, spin3}. Unlike ordinary orders, there is no symmetry breaking mechanism \cite{sym} for understanding a topological order. In fact, there are different topological phases with the same symmetry which clearly violates the Landau paradigm of phase transitions \cite{wen, wen2}. A physical system with topological order shows some interesting and exotic properties which do not have any analogue in other physical systems. Specially, topological degeneracy of the ground state of such systems is robust against local perturbations so that these states are important candidate for robust storage of quantum information\cite{TC}. Furthermore, excitations of the topological models are quasi-particles with exotic statistics, called anyons \cite{wen3, wen4}. Braiding an anyon around another one leads to multiplication a phase factor $e^{i\theta}$ for abelian anyons or an unitary operator for non abelian anyons to their wave functions.\\

Specifically, there are two important sets of topological codes namely Kitaev's toric code states (TC) and color code states (CC) that we generally call CSS topological states. The TC are simple models with $Z_d$ topological order and are defined on oriented graphs \cite{TC}, for $Z_2$ models the graphs can be non-oriented. They are defined on 2D lattices with different number of holes or different topological structures, called surface codes \cite{sur}, and also on 3D lattices with different topological properties \cite{3D, 3Dt}. The CC states are also other simple models with $Z_2 \times Z_2$ topological order \cite{cc}. They are ordinary defined on three-colorable lattices, technically called colexes and can be generalized to arbitrary dimensions \cite{bo}. It has been shown that a CC on 3-colexes can be used for a universal quantum computation \cite{ccu}.\\

In spite of exciting theoretical properties of topological models, there are some important challenges for experimental realization of anyons. Specifically direct observation and control of anyons is hard in most physical systems \cite{ex, cirac} so that only simple topological objects are accessible. The experimental realization of anyons will be the more challenging if one tries to generate non-abelian anyons which are more important due to ability of universal quantum computation \cite{fr}. An alternative way to overcome such challenges are the extrinsic defects \cite{2, 3, 4, 5, 6, 7, 8, 9, 10, 11, 12}. The extrinsic defects are point-like objects which are not finite-energy excitations of a physical system. In lattice models, the extrinsic defects are usually created by dislocations in the lattice. There is also a proposal for creating and manipulating them by an external field \cite{mag}.\\

In this paper we concentrate on a specific method of generating the topological states that we call measurement-based preparation. Since the topological order is a long range entanglement in a quantum state, it is not necessary to generate a Hamiltonian for a topological state but one can generate a topological state by a more direct approach. Specially measurement-based quantum computation \cite{m1, m2, m3, m4, m5, m6, m7, m8, m9} is a candidate for this goal \cite{fo}. In \cite{ros0}, authors gave a simple pattern for generating a TC state on a square lattice by single-qubit measurements on a 2D cluster state. Therefore, one can experimentally realize a TC state and also charge and flux anyons just by single-qubit measurements and operations on a 2D cluster state \cite{ros1}. \\

The extension of the above idea to other CSS topological states is the main goal of this paper. While in the previous works, only a TC state on a simple 2D lattice had been considered, here we give a general scheme for determining measurement patterns corresponding to different CSS topological states. We explicitly consider TC states and CC states on different lattices in various dimensions and different topological structures. We give a set of step-by-step transformations which are supported by graphical notations to find the measurement patterns of the above topological states. In order to show how our scheme works, we use it for two specific examples, TC on a 3D lattice and CC on a 3-colex. We emphasize that although all quantum states can be generated by single-qubit measurements on the cluster states, it is not a simple task to find different measurement patterns for generating different quantum states. Furthermore, since cluster states can be realized by optical lattices \cite{opt}, our scheme can be used for experimental realization of topological states on cluster states. \\

Furthermore, we study measurement-induced defects on the toric code states. We show that some specific measurement patterns lead to generating the topological states with the richer structures. Specifically, we give a measurement-based scheme for generating Ising anyons. Since Ising anyons are non-abelian anyons, they have richer topological properties than abelian case. We show that a measurement-induced defect plays the role of a twist in the toric code state similar to one introduced in \cite{9} where author has shown that a twist has all the properties of an Ising anyon. Therefore our scheme will be a new method for generating Ising anyons.\\

The rest of this paper is as follows: In section (\ref{s1}), we review three important classes of quantum states containing graph states, TC states and CC states. In section (\ref{s2}), We give a method for generating the CSS states containing the TC states and the CC states by single-qubit measurements on the cluster states. To this end, firstly we introduce a pattern for generating the CSS topological states on some complex graph states. In the second step we givea few graphical rules for generating the above graph states from a cluster state. These two steps lead to a general scheme for measurement-based preparation of CSS topological states on the cluster states. In section (\ref{ss})We will also give two three-dimensional examples to show how our scheme works. In section (\ref{s3}), we also propose a measurement pattern for generating non-abelian anyons, the Ising anyons. To this end, we study single-qubit measurements in the Pauli base $Y$ on a toric code state. We show that it lead to generating topological defects which play role of Ising anyons. \\
\section{Three classes of important quantum states}\label{s1}
In this section we review three classes of quantum states namely graph states, TC states and CC states. While graph states are known as resources for measurement-based quantum computation, TC states and CC states are states with topological order which are specifically important for topological quantum computation.
\subsection{Graph states}
Consider an arbitrary graph $G=(V,E)$ where $V$ is the set of
vertices and $E$ is the set of edges which connect different
vertices of the graph. Corresponding to each graph, we can define
a quantum state which is called graph state. To this end, suppose
that there are qubits which live on vertices of the graph and we
have prepared the state of each qubit in the quantum state $|+\ra$
which is the positive eigenstate of the Pauli operator $\sigma_x$, we denote this operator by $X$.
Then we consider a two-qubit operator which is called $CZ$ in the
following form:
\begin{equation}
C\mathbb{Z}=|0\ra \la 0| \otimes I + |1\ra \la 1| \otimes Z
\end{equation}
where $Z$ refers to the Pauli operator $\sigma_z$ and $|0\ra$ and
$|1\ra$ are positive and negative eigenstates of $Z$ where $Z|0\ra=|0\ra$ and $Z|1\ra=-|1\ra$. If we apply
such operators on all pairs of qubits which are neighbor in the graph,
we will generate a graph state. Therefore, a graph state $|G\ra$
can be written in the following simple form:
\begin{figure}[t]
\centering
\includegraphics[width=8cm,height=8cm,angle=0]{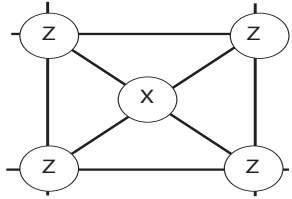}
\caption{(Color online) A stabilizer corresponding to central qubit on a graph state with five qubits } \label{gstate}
\end{figure}
\begin{equation}\label{x1}
|G\ra= \prod_{e}CZ_e |+\ra^{\otimes |V|},
\end{equation}
where $e$ refers to an edge of the graph and $CZ_e$ is a
$CZ$ operator which is applied on two qubits in two endpoints of
the edge $e$. Graph states can also be defined by their stabilizers group. A stabilizer group $S$ is an abelian subgroup of the Pauli group on n qubits which excludes the element $-I$. The state which is stabilized by all the elements of $S$, is called the stabilizer state. It is simple to show that a graph state in the form of (\ref{x1}) is stabilized by a group of Pauli operators in
the following form, see Figure (\ref{gstate}):
\begin{equation}
S_a =X_a \prod_{b\in N_{a}}Z_b
\end{equation}
Where we have defined the operator $S_a$ corresponding to each vertex $a$ of the graph and $N_a$ is the set of neighbors of the vertex $a$.\\

Graph states have the important property that all unitary operators on them can be induced by appropriate sequence of single-qubit measurements on their vertices \cite{m1, m2}. It has been shown that a graph state corresponding to a one-dimensional graph is not a universal resource for quantum computation. A minimal requirement for universal quantum computation is a graph state on a two-dimensional square lattice which is called the cluster state. It has been shown that each unitary operator can be realized by single-qubit measurements on an enlarged cluster state. In other words, any multi qubit state can be prepared on the remaining set of vertices where a specific set of vertices are measured in appropriate bases.\\

\subsection{Kitaev's toric code states}
Kitaev's toric code states (TC), are an important example of CSS topological states with
$Z_2$ topological order. We call them CSS states because their stabilizers are in the form of $X$-type and $Z$-type operators. TC can be defined on arbitrary graphs where qubits live in edges of the graph, as an example we consider
a simple square lattice with qubits on edges, as it is shown in Figure (\ref{toric}). We
define two $X$-type and $Z$-type commutative operators corresponding to each plaquette
and vertex of the lattice in the following form:
\begin{figure}[t]
\centering
\includegraphics[width=7cm,height=9cm,angle=0]{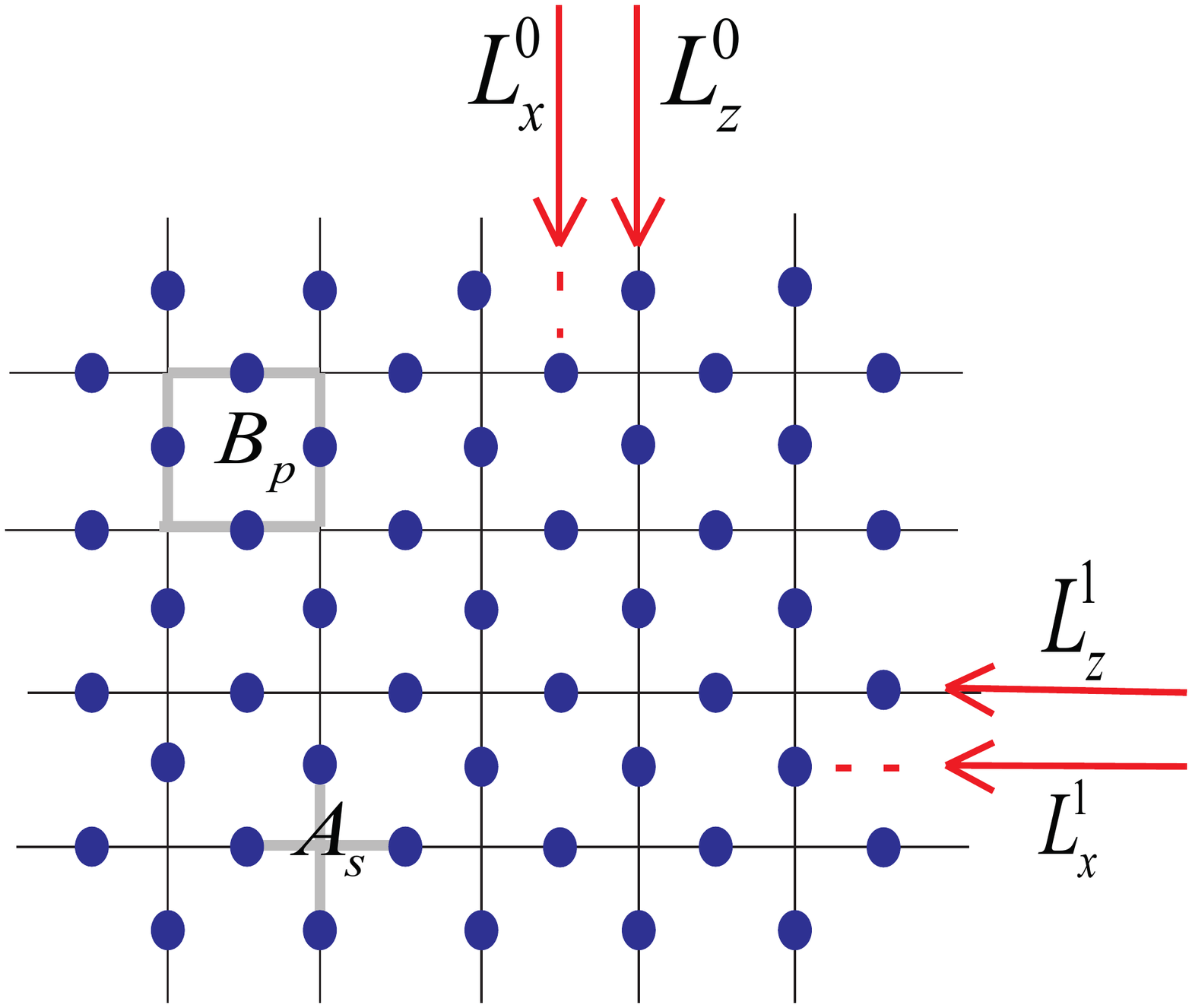}
\caption{(Color online) The plaquette and vertex operators $A_s$ and $B_p$ of the TC state on a square lattice. There are two non-contractible loops corresponding to each directions of the lattice on a torus. For each direction $\sigma=\{0,1\}$, two loop operators $L_{x}^{\sigma}$ and $L_{z}^{\sigma}$ is constructed by $Z$ and $X$ operators.} \label{toric}
\end{figure}
\begin{equation}\label{x2}
B_p =\prod_{i\in \partial P}Z_i~~~,~~~A_s=\prod_{i\in s} X_i
\end{equation}
where $i\in \partial p$ refers to edges around a plaquette and $i\in s$ refers to edges pointing to a vertex and $Z$ and $X$ are the Pauli operators. Furthermore, according to the boundary conditions of the lattice, there are two constraints on the plaquette and vertex operators in the form of $\prod_{p}B_p =1$ and $\prod_{v}A_v =1$. It shows that all plaquette and vertex operators are not independent. consequently, we can only consider the independent stabilizers of the model to find the corresponding stabilizer state. To this end, since vertex and plaquette operators commute with each other, it is simple to show that the following state is a stabilizer state of the above operators (\ref{x2}):
\begin{equation}\label{x4}
|\psi \rangle = \frac{1}{\sqrt{2}^N}\prod_{p\in } (1+B_p)|+++...+\rangle,
\end{equation}
where $\prod_{p}$ refers to product of all plaquettes of the lattice whose corresponding operators are independent and $N$ is the number of independent plaquettes. the product state of $|+++...+\ra$ is defined on qubits living on all edges of the graph. Such a state has a topological order which leads to a robust degeneracy in the stabilizer state of the model \cite{karimipour, zarei, kargarian}. This property can be well found when we define the model on a torus. In fact, on a torus topology, there are also non-contractible loop operators in the form of:
\begin{equation}
L_{z}^{\sigma}=\prod_{i\in L^{\sigma}}Z_i
\end{equation}
where $i\in L^{\sigma}$ refers to the qubits which live on a non-contractible loop around the torus in two different directions $\sigma=0~or~1$, see Figure (\ref{toric}). In this way, the following four quantum states are the stabilizer states of the operators (\ref{x2}).
\begin{equation}\label{x3}
|\psi_{i,j}\ra=(L_{z}^{0})^i (L_{z}^{1})^j |\psi\rangle
\end{equation}
where $i,j$ have values $0$ or $1$ corresponding to four different quantum states.\\

Topological order of the TC states are understood by three important properties, robust degeneracy, non-local order parameter and anyonic excitations. The robust degeneracy is due to the fact that four degenerate stabilizer states (\ref{x3}) can not be distinguished from each other or converted to each other by any local parameters so the degeneracy is robust against all local perturbations. In fact, there are two non-local operators which have different expectation values in different degenerate stabilizer states. Such operators are defined in the following form:
\begin{equation}
L_{x}^{\sigma}=\prod_{i\in L^{\sigma}}X_i
\end{equation}
where $i\in L^{\sigma}$ refers to the qubits which live on a non-contractible loop around the torus in two different directions $\sigma=0~or~1$ and $X$ is the Pauli operator, see Figure (\ref{toric}). The expectation values of these operators in each one of the four degenerate stabilizer states are as follows:
$$ \la \psi_{00}|L_{x}^{0}|\psi_{00}\ra =1~~,~~\la \psi_{00}|L_{x}^{1}|\psi_{00}\ra =1$$
$$ \la \psi_{01}|L_{x}^{0}|\psi_{01}\ra =-1~~,~~\la \psi_{01}|L_{x}^{1}|\psi_{01}\ra =1$$
$$ \la \psi_{10}|L_{x}^{0}|\psi_{10}\ra =1~~,~~\la \psi_{10}|L_{x}^{1}|\psi_{10}\ra =-1$$
\begin{equation}
\la \psi_{11}|L_{x}^{0}|\psi_{11}\ra =-1~~,~~\la \psi_{11}|L_{x}^{1}|\psi_{11}\ra =-1
\end{equation}
Therefore these two non-local operators can distinguish the different stabilizer states.\\

\begin{figure}[t]
\centering
\includegraphics[width=9cm,height=9cm,angle=0]{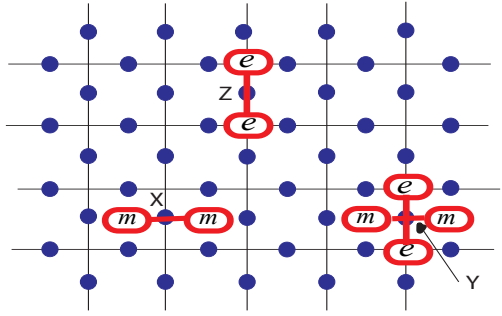}
\caption{(Color online) A pair of charge (flux) anyons is created by applying an operator $X$ ($Z$) on one of the qubits. A pair of $\epsilon=e\times m$ anyons is created by applying an operator $Y$ on a qubit.} \label{epsi}
\end{figure}
\begin{figure}[t]
\centering
\includegraphics[width=9cm,height=9cm,angle=0]{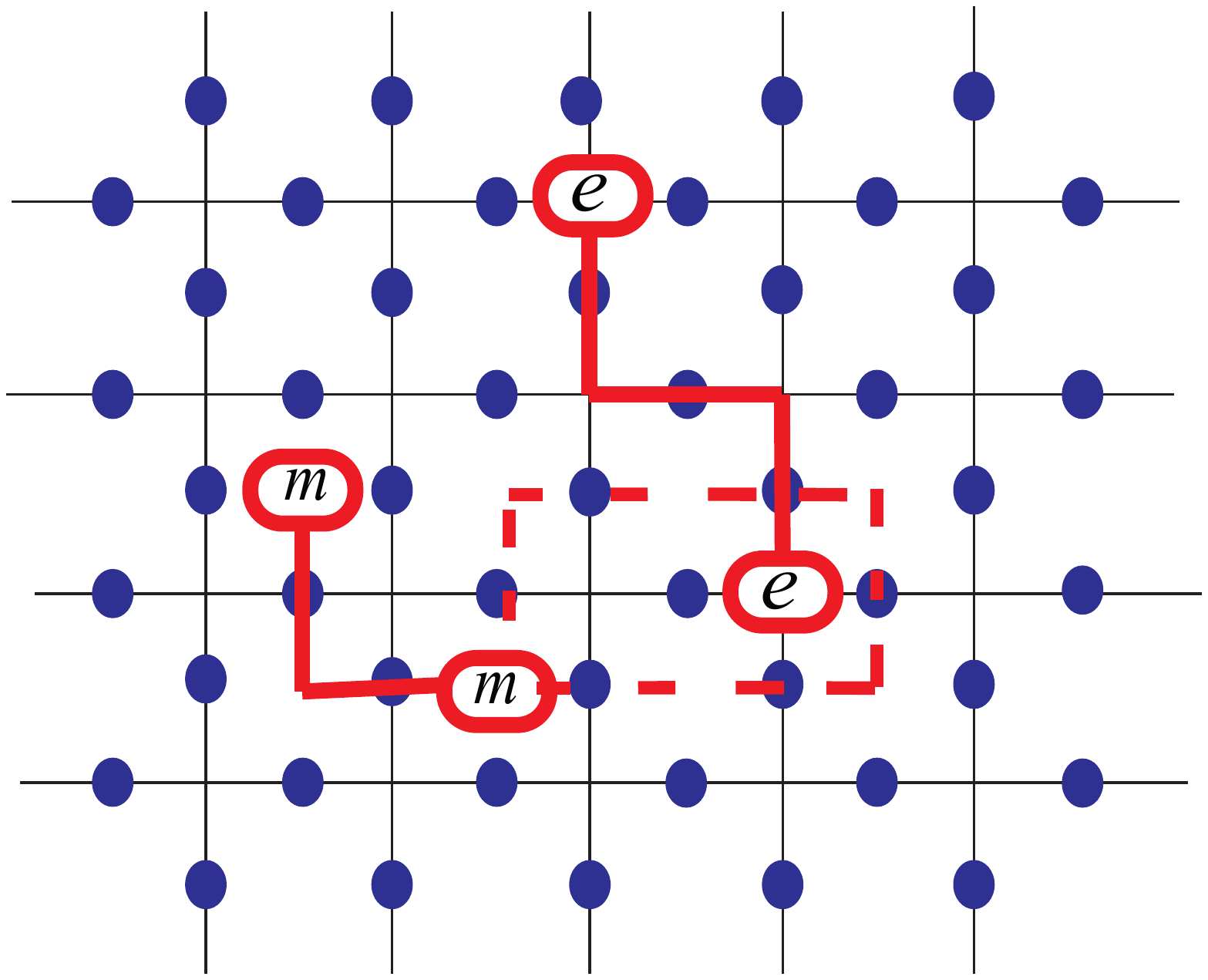}
\caption{(Color online) A pair of charge (flux) anyons is created by applying a operator $X$ ($Z$) on one of the qubits. A charge (flux) anyon can move on the lattice (dual lattice) with applying a string of operators $X$. Dotted line shows the winding a flux anyon around a charge anyon, it leads to a minus sign to the wave function.} \label{anyons}
\end{figure}
Finally, excitations of the TC model are quasi-particles with anyonic statistics. They make an anyonic model with four particles, vacuum $1$, charge anyon $e$, flux anyon $m$ and fermion $\epsilon$. The fusion rule of such particles is as follows:
$$ x\times x=1~, x\times 1=x,~~~where~x=\{1,e,m,\epsilon \}$$
\begin{equation}
e\times m=\epsilon~,~e\times \epsilon =m~,~m\times \epsilon =e
\end{equation}
The excitations are generated by applying the Pauli operators $X$ or $Z$ or $Y$ on a qubit of the lattice. An $X$ operator on a qubit does not commute with two plaquette operators which are share in that qubit and it is interpreted as two flux anyons $m$ in two corresponding plaquettes, see Figure (\ref{epsi}). Also a $Z$ operator does not commute with two neighbor vertex operators and it is interpreted as two charge anyons $e$ in two corresponding vertices. Since $Y=iXZ$, this operator generates two charge particles and two flux particles on the corresponding vertex and plaquettes which is equal to two fermions $\epsilon$, see Figure (\ref{epsi}). By applying a string of $X$ ($Z$) operators, a flux (charge) anyon moves on plaquettes (vertices) of the lattice, see Figure (\ref{anyons}). It is simple to show that if a charge anyon winds around a flux anyon, it leads to a minus sign on the wave function, see Figure (\ref{anyons}). Such a factor shows that charge and flux anyons are not fermions or bosons. \\

\begin{figure}[t]
\centering
\includegraphics[width=7cm,height=7cm,angle=0]{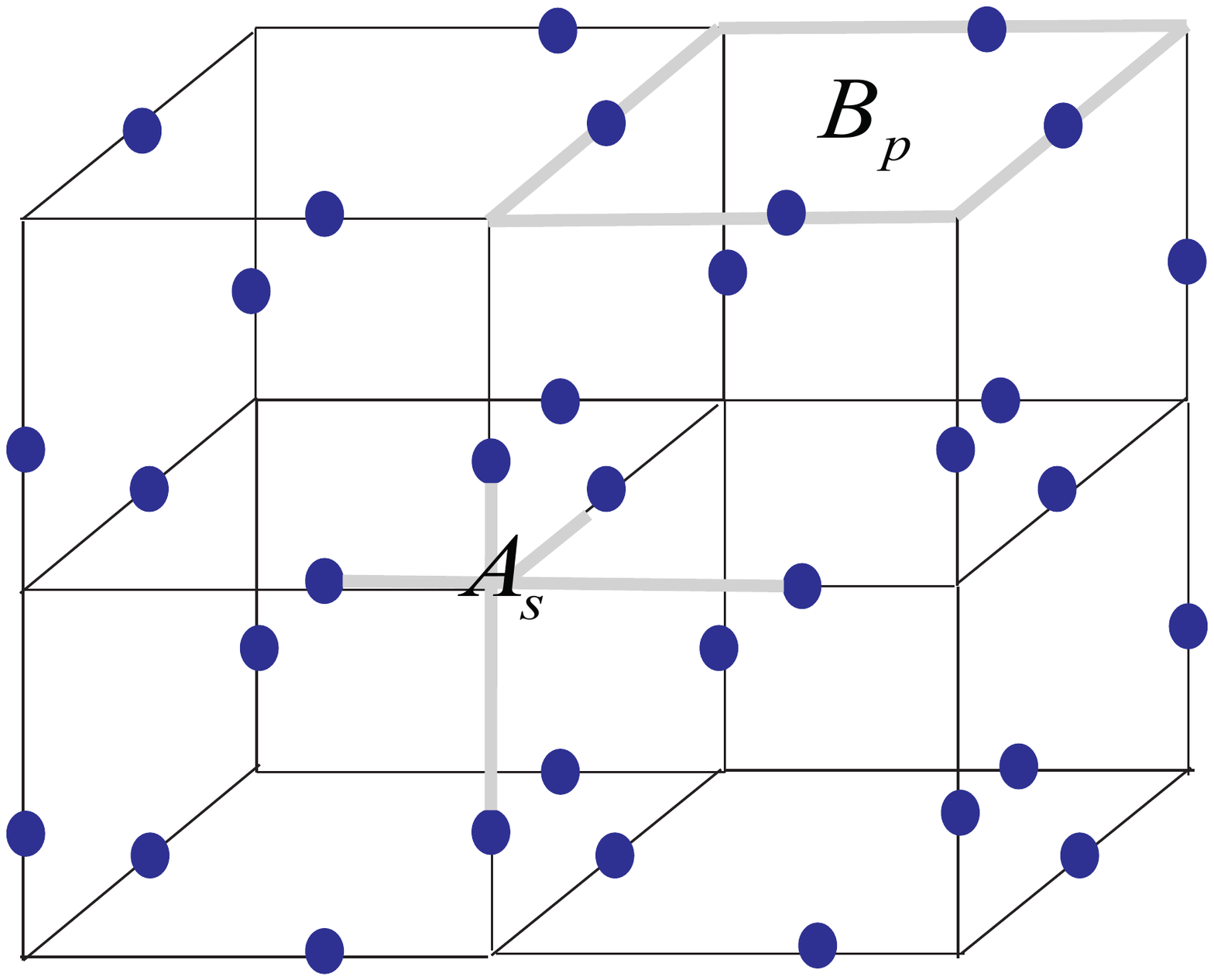}
\caption{(Color online) TC on a 3D lattice where qubits live in edges of the lattice. $Z$-type operators are defined corresponding to each plaquette of the lattice and $X$-type operators are defined corresponding to each vertices of the lattice.} \label{3Dto}
\end{figure}
Although we use the TC model on the torus to explain topological order of that model, the TC states can be defined on lattices with different topologies. For example, we can generate a hole in an arbitrary two-dimensional lattice by removing one of the plaquette operators from stabilizers (\ref{x2}). In this way one can increase degeneracy of the model by adding the more holes in the lattice. Furthermore, TC states can be defined on the lattices in the higher dimensions. For example, in Figure (\ref{3Dto}) we show a TC on a 3D square lattice where other topological properties can be realized \cite{3D}.
\subsection{CC states}
\begin{figure}[t]
\centering
\includegraphics[width=7cm,height=7cm,angle=0]{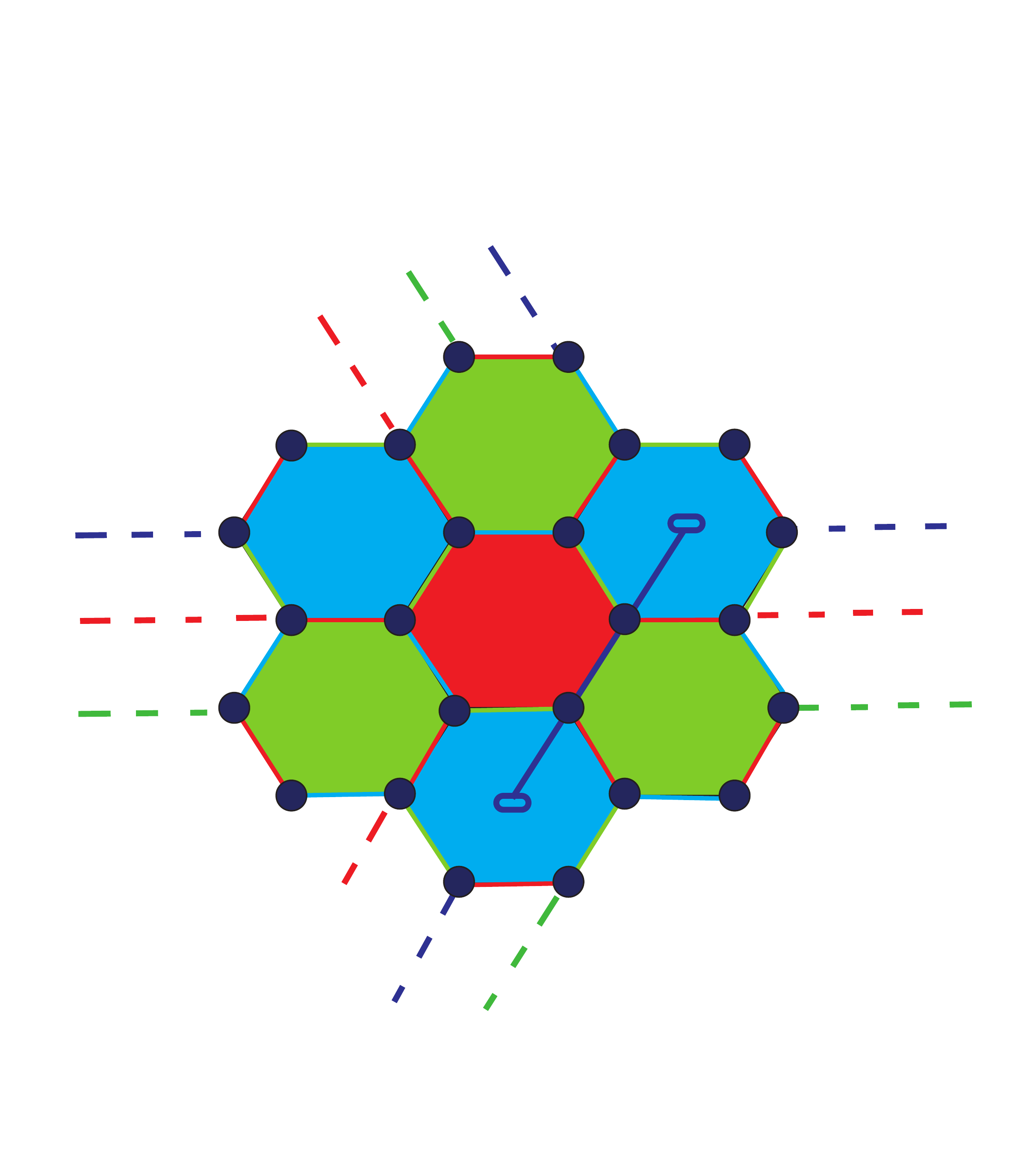}
\caption{(Color online) Plaquettes of a hexagonal lattice are colored by three different colors. There are three non-contractible loops in two directions of the lattice corresponding to three colors of the plaquettes. A string of $X$ or $Z$ operators can generates two excitation in two plaquettes corresponding to two end-points of that string.} \label{color}
\end{figure}
Another well-known CSS topological state is the CC with
$Z_2 \times Z_2$ topological order. The CC models can be defined on 2D
three-colorable lattices which technically are called 2-colexes and can
be generalized to arbitrary dimensions \cite{bo}. In figure
(\ref{color}), we show a simple example of such a model on a hexagonal
lattice. Qubits live on the vertices of the lattice and two
stabilizers are defined corresponding to any plaquettes of the
lattice as follows:
\begin{equation}
h_x =\prod_{i\in \partial h} X_i~~,~~h_z =\prod_{i\in \partial h} Z_i
\end{equation}
where $i \in \partial h$ refers to all qubits belonging to the boundary of the plaquette $h$. Since the above operators are members of the Pauli group and they commute with each other, it is simple to show that the following state is a stabilizer state of the model:
\begin{equation}\label{c1}
|\phi_c \ra =\frac{1}{\sqrt{2}^N}\prod_{h}(1+h_z )|++...+\rangle
\end{equation}
where $\prod_{h}$ refers to product of all plaquettes of the lattice whose corresponding operators are independent. Moreover, $N$ is the number of the independent plaquettes of the lattice and is determined by the boundary conditions of the lattice . Although the form of this state is very similar to the TC states in (\ref{x4}), there is an important difference with the TC model which leads to richer topological structure of the CC model. In fact, the degree of freedom of color in the CC leads to a higher degree of degeneracy than the TC model \cite{zarei2}. As it is shown in Figure (\ref{color}), there are three non-contractible loops in each direction of the lattice corresponding to three colors of the plaquettes. In fact the excitations of CC model can be generated in plaquettes with three different colors and each excitation with a specific color can be moved in the plaquettes with the same colors by applying a string of $X$ or $Z$ operators on edges with the same color, see Figure (\ref{color}).\\ 

Furthermore, It has also been shown that the CC states are more efficient than the TC states for quantum computation where there is also possibility of universal quantum computation with the CC models on a 3D lattices \cite{ccu}. The CC states have been also extended to higher dimensions where the structure of the model is defined based on D-colexes in a D-manifold \cite{bo}. In spite of very different properties of such models, the form of stabilizers of them is similar to the 2D case as $X$-type and $Z$-type operators in the following form:
\begin{figure}[t]
\centering
\includegraphics[width=12cm,height=10cm,angle=0]{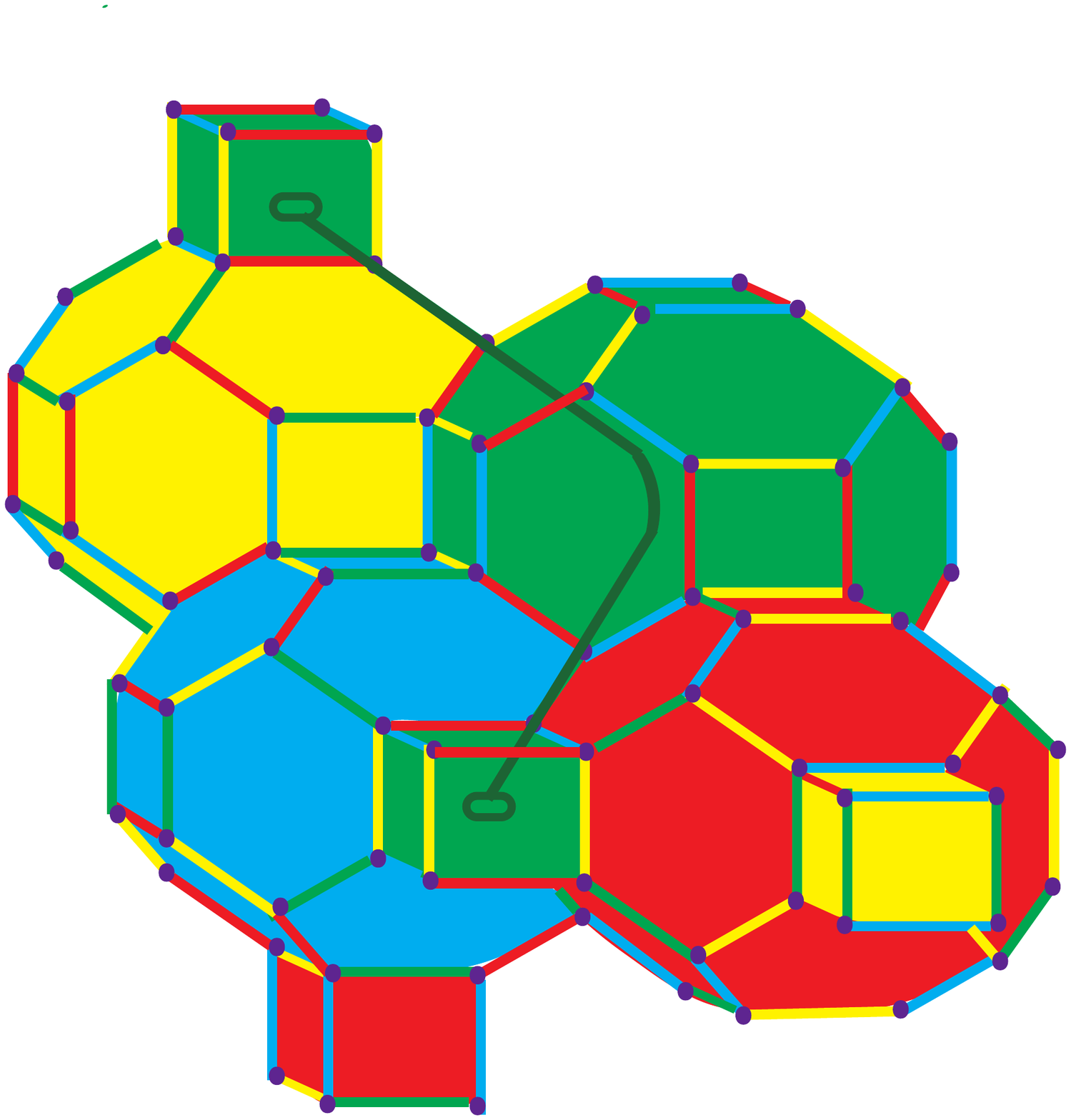}
\caption{(Color online) CC on a 3-colex where qubits live in the vertices of the lattice. All cells and edges of the lattice are colored by four different colors where an edge with a specific color connects two cells with the same color. $X$-type operators are defined corresponding to each cell of the lattice and $Z$-type operators are defined corresponding to each plaquette of the lattice. a string of $Z$ operators on edges with a specific color generates excitations in two cells corresponding to end-points of that string. } \label{3Dcolor}
\end{figure}
\begin{equation}
C_z =\prod_{i\in c}Z_i~~,~~C'_x = \prod_{i\in c'}X_i
\end{equation}
where $c$ and $c'$ denote those cells of a D-colex whose corresponding operators commute with each other. By such a definition the ground state of this model would be written in a similar form of (\ref{c1}) as follows:
\begin{equation}\label{c10}
|\phi_D \ra =\frac{1}{\sqrt{2}^M}\prod_{c}(1+C_z )|++...+\rangle
\end{equation}
where $\prod_{c}$ refers to product of all cells corresponding to the independent $Z$-type operators. Moreover, $M$ is the number of independent $Z$-type cells and is determined by boundary conditions of the lattice. For example in Figure (\ref{3Dcolor}), we show a
3-colex where all cells and edges are colored by four different
colors and each cell with a specific color has involved by edges
with other three colors. The $Z$-type operators are defined
corresponding to the plaquettes of the lattice and the $X$-type
operators are defined corresponding to the cells of the lattice.
Similar to the 2D model, the colorability of the 3-coloex plays an
important role for describing excitations of the model. Specially,
cells of the lattice with a specific color are connected to each
other by edges with the same color. A pair of excitations can be
generated into two different cells with the same color by applying
a string of $Z$ operators on successive edges between them, see
Figure (\ref{3Dcolor}). Another kind of excitations can be
generated by applying $X$ operators, but they are shown by
membranes instead of strings \cite{bo}.
\section{Preparation of CSS topological states from 2D cluster states}\label{s2}
As we mentioned in the previous section, cluster states are universal resources for quantum computation where any arbitrary quantum states can be produced by single-qubit measurements on cluster states. In spite of such an important property for cluster states, it is not a simple task to explicitly find which kinds of single-qubit measurements can generate a specific quantum state. In fact as a general method one should find a unitary evolution corresponding to generating a quantum state. Then the unitary operator can be decomposed to the $CZ$ gates and single-qubit operators. Finally by a measurement pattern for single-qubit operators and $CZ$ gates, one will be able to generate any quantum states starting from a cluster state.\\

Finding measurement pattern of different quantum states can be easier if one wants to generate a stabilizer state. Since the cluster states are also stabilizer states, it is shown that one can produce a stabilizer state by single-qubit measurements in the Pauli bases on the cluster states. Specially the CSS topological states like TC and CC states belong to the set of stabilizer states and the Pauli bases are enough for measurement-based preparation of them. \\

Although some important attempts for measurement-based preparing some specific topological models has been done \cite{fo, ros0, ros1}, here we give a more general and systematic scheme for generating different CSS topological states from 2D cluster states. In this section, we show how TC states and CC states on different lattices with various topological structures and in various dimensions can be produced by single-qubit measurements on the cluster states. In this way we give a general scheme for preparing CSS topological states by measurement-based quantum computation. By a finite set of graphical transformations, we will explicitly find measurement pattern corresponding to each CSS topological state.\\

To this end, we firstly show any the above CSS topological states can be generated by measurements in the $X$-basis on a bipartite graph state. Then we show that the bipartite graph states can be embedded on a plan as a non-planar graph. In the next step, we give a measurement rule to transform the non-planar graph to a planar graph. Finally, we will give other measurement rules to transform the planar graph to a 2D square lattice corresponding to a 2D cluster state. Furthermore, we will also give a set of graphical transformations corresponding to the above measurement rules to find the measurement pattern corresponding to various CSS topological states on a cluster state.
\subsection{Transformation to a bipartite graph state}
In this subsection, we show how a CSS topological state can be
generated by single-qubit measurements on a bipartite graph state.
we call such a transformation a cell insertion. In a recent paper
\cite{fo}, a similar idea has been used for generating the CSS
quantum states from a 3D cluster state. To this end, we consider a
general form of a CSS state in the following form:
\begin{figure}[t]
\centering
\includegraphics[width=12cm,height=12cm,angle=0]{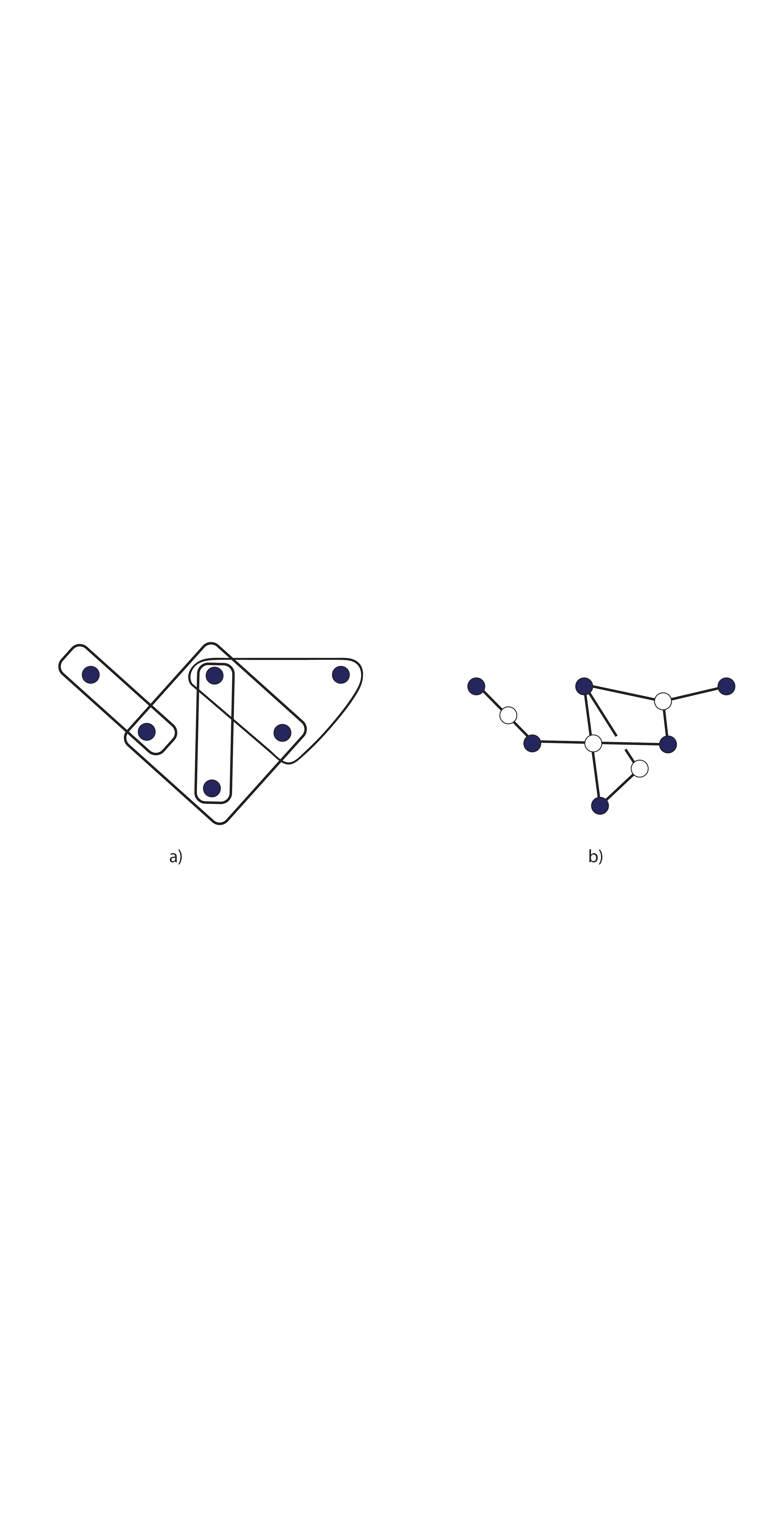}
\caption{(Color online), in the left hand, we show a CSS quantum state where each closed curve show a $z$-cell of the model. In the left hand, we show a bipartite graph state corresponding to the CSS state in the left hand. White qubits should be measured in the $X$-basis.} \label{sexam}
\end{figure}
\begin{equation}\label{q0}
|\psi\ra =\frac{1}{\sqrt{2}^M}\prod_{c} (1+C_z) |+++...+\rangle
\end{equation}
where $\prod_{c}$ refers to product of all cells corresponding to the independent $Z$-type operators of the CSS topological state. Moreover, $M$ is the number of the independent set of $Z$-type operators and is determined by boundary conditions of the lattice. Furthermore, $C_z =\prod_{i\in c}Z_i$ is a product of the Pauli operators $Z$ on qubits belonging to the cell $c$, we call them $z$-cells. It is clear that all TC and CC states that we defined in the previous section are a kind of the above state. In Figure (\ref{sexam}-a), we show a simple example of a CSS state where qubits are denoted by black circles and each $z$-cell $c$ is denoted by a closed curve around the corresponding qubits.\\

In order to find a bipartite graph corresponding to the initial model, we add a new qubit which is denoted by a white circle corresponding to each cell of the initial model. If we connect the white qubit corresponding to each cell to the black qubits belonging to the same cell, we will have a bipartite graph, see Figure (\ref{sexam}-b). We define a graph state on the above bipartite graph and show that measurement of white qubits of the bipartite graph state in the $X$-basis leads to the initial CSS state on the black qubits. We can understand it by considering stabilizers of the bipartite graph state. To this end, consider a stabilizer related to a white qubit $w$ in the form of $X_w  \prod_{b\in N(w)}Z_b$ where $N(w)$ refers to black qubits neighboring to the white qubit $w$. It is clear when we measure the qubit $w$ in the $X$-basis, an operator as $\prod_{b\in N(w)}Z_b$ remains on the black qubits and it is the $Z$-type operator of the initial model. By proper multiplication of stabilizers corresponding to the black qubits of the bipartite graph state, we will also find the $X$-type operators. This argument hold for any CSS states in the form of (\ref{q0}) and we give an explicit proof of this statement in appendix A.\\

There is also an important problem where the result of measurement of a white qubit in the $X$-basis can be $-1$ and therefore the corresponding $Z$-type operator after measurement will be in the form of $-\prod_{b\in N(w)}Z_b$, we call such a cell of the initial model an excited cell. We should show that such probabilistic errors can be corrected by applying successive $X$ operators. To this end, we save the result of all measurements till the end of the process. The final CSS topological state will be an excited state of the original model where there are excitations on the excited cells. Method of correcting this errors depends on the structure of topological state specially the nature of excitations of the initial model. For example, for TC on a 2D lattice a string of $X$ operators can be applied between a pair of the excited cells which leads to correcting the corresponding errors. Furthermore, we can also apply a string of $X$ operators from one excited cell to the boarder of the lattice. There is also a similar method for CC on 2D lattice where one can apply a string of $X$ operators to correct errors. But there is a difference that each string can be constructed by a specific color which moves on edges with the same color. Therefore the excited cells of the model with a similar color can be corrected by a string between themselves. In higher dimension the method of error correction should be changed corresponding to the structure of the topological model. In the next section, we give also two 3D examples to show how our scheme can work.\\

In this way, we showed that all CSS topological states can be generated by single-qubit measurements on a bipartite graph state. In the next subsection we give a few measurement rules, namely uniformizing and flattening rules, to generate the bipartite graph states from the 2D cluster states. In \cite{grstate}, It has been shown that Pauli measurements on a graph state lead to a new graph state on another graph by a few graphical transformations. We use these facts to find our measurement rules in the next subsection.\\

\subsection{Uniformizing the graph}
As we showed, in order to generate a CSS topological state from a bipartite graph state, corresponding to each z-cell of the original model we have to add a new vertex which is connected to all vertices taking part in that z-cell. This leads to a bipartite graph with vertices with various degrees of connectivity. Here we give a rule to uniformize the graph by reducing degrees of all vertices to four. We call this rule a uniformizing rule.\\

\textbf{Uniformizing rule:}\\

\begin{figure}[t]
\centering
\includegraphics[width=8cm,height=6cm,angle=0]{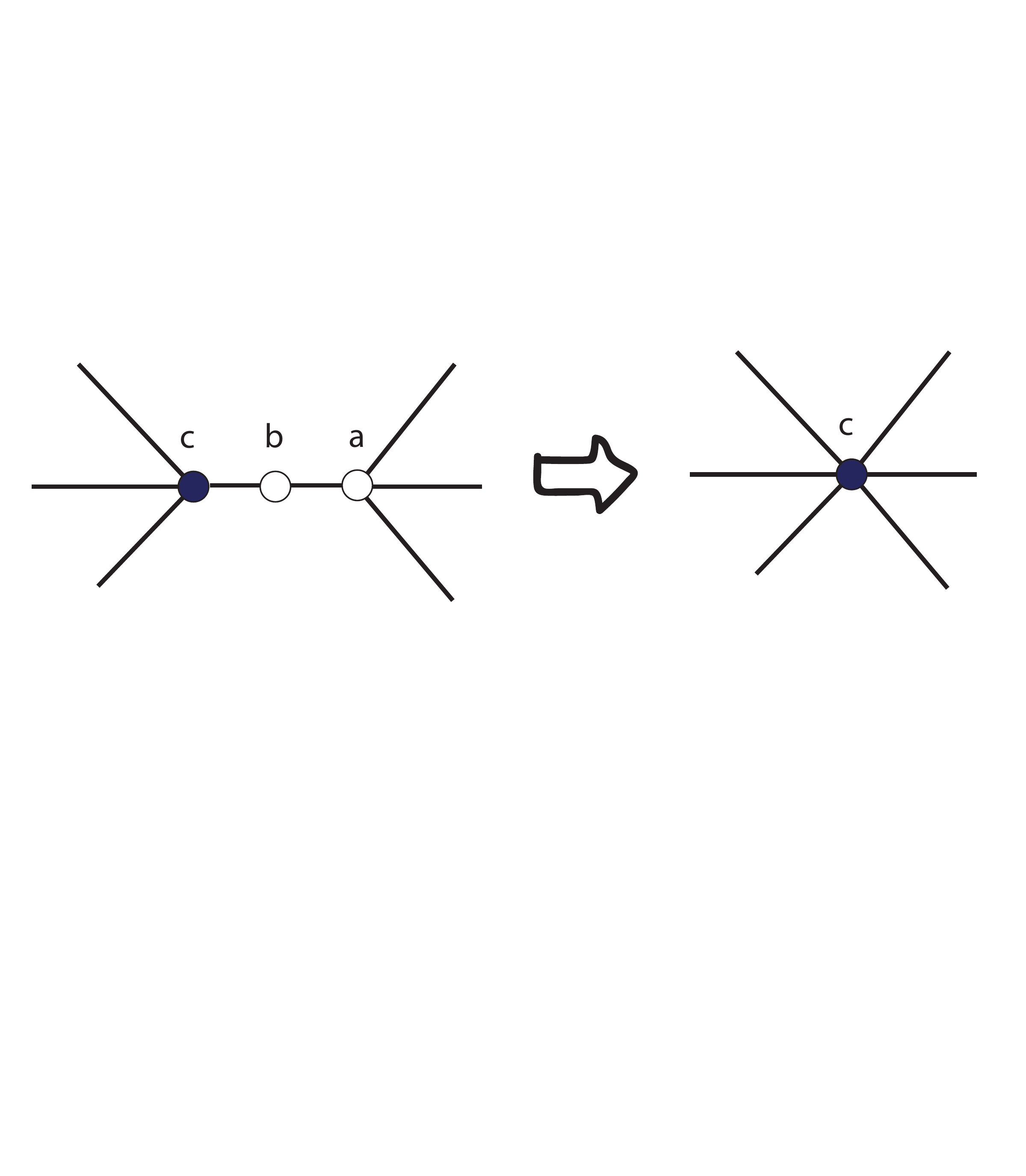}
\caption{(Color online) In the left hand, there is a graph state where the white qubits are measured in the $X$-basis. It leads to a new graph state on unmeasured qubits as shown the right hand. where two vertices of the initial graph are merged to each other.} \label{merging}
\end{figure}

Consider a graph state corresponding to the graph that has been shown in Figure (\ref{merging}, left). We measure firstly qubit "a" and then "b" which are denoted by white circles in the $X$-basis. After measurement, those stabilizers of the graph state that do not commute with an $X$ operator on $a$ and $b$ should be removed and the graph state is converted to another one on a new graph where two vertices of initial graph have been merged to each other. Since the result of each measurement is probabilistic, after each measurement a correction should be applied as follows. \\

If the result of measurement on qubit $a$ was $-1$, we would apply an operator in the form of $X_b Z_c$ to correct the error. After measurement of the qubit $b$, if the result was $-1$, we would apply an operator in the following form to correct the error:
\begin{equation}
X_c \prod_{i\in N_{c}}Z_i
\end{equation}
where $i\in N(c)$ refers to all qubits neighboring to the qubit $c$ except of $b$. \\

In this way, we can reduce the degree of each vertex of the graph to four. Furthermore, since we finally want to transform the graph to a 2D rectangular lattice, we should also add new vertices to already existing links or plaquettes. To this end we also give two other rules as follows:\\

\textbf{face insertion:}\\

\begin{figure}[t]
\centering
\includegraphics[width=10cm,height=7cm,angle=0]{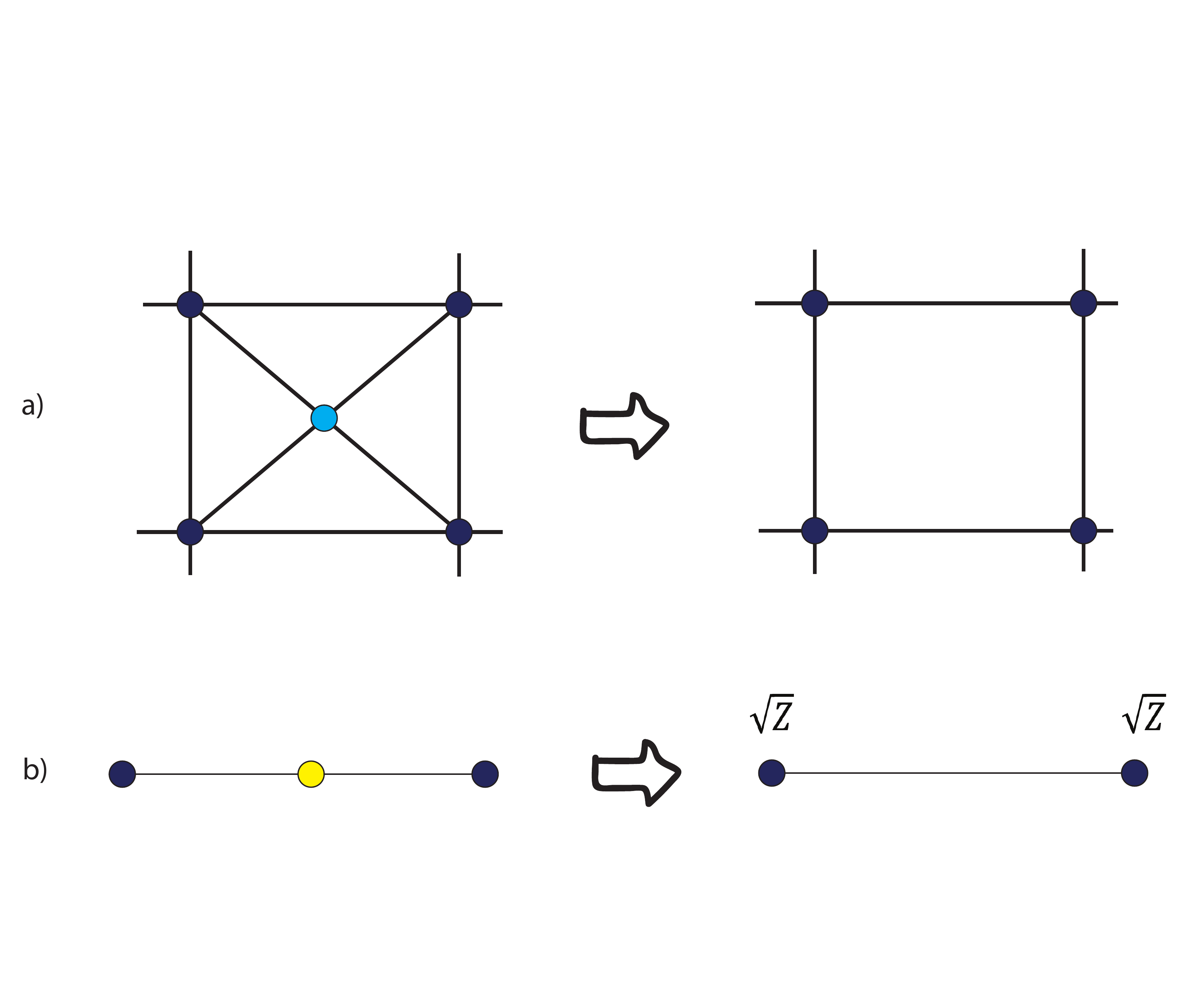}
\caption{(Color online) a) In the left hand, there is a graph state where the blue qubit is measured in the $Z$-basis. In the right hand, there is a new graph state on unmeasured qubits. where the blue qubit of the initial graph is removed. b) in the left hand there is a graph state where the yellow qubit is measured in the $Y$-basis. It leads to a new graph state on unmeasured qubits as shown in the right hand. Furthermore, After measurement of the yellow qubit in the left hand, it is necessary to apply an operator $\sqrt{Z}$ on unmeasured qubits to generate a graph state. } \label{removing}
\end{figure}

This rule is related to the measurement in the $Z$-basis. In Figure (\ref{removing}-a), we show that if one measures the qubit "a" of a graph state which is denoted by blue circle in the $Z$-basis, it will be converted to a new graph state where the measured qubit has been removed. Similar to the previous rule, studying commutation of the stabilizers of the graph state with an operator $Z$ on the qubit $a$ will lead to the proof of this rule. There is also a probabilistic error where the result of the measurement can be $-1$. However, it is enough to apply an operator as $\prod_{i\in N(a)}Z_i$ to correct the error where $N(a)$ refers to all neighbors of the qubit $a$.\\

\textbf{Link insertion:}\\

As it is shown in Figure (\ref{removing}-b), we can also add a new qubit on a link between two vertices. If we measure the new qubit in the $Y$-basis, we will have the original graph. Furthermore, it is necessary to apply an operator $\sqrt{Z}$ on the unmeasured qubits. In order to correct a probabilistic error, it is also enough to apply the operator $Z$ on both neighbors of the measured qubit.\\

\subsection{Flattening the graph}
Depending on the original CSS topological state, the corresponding bipartite graph can be also a non-planar graph and uniformizing degrees of the vertices of the graph can not remove non-planarity of the graph. Here we give another rule to solve this problem.\\

\textbf{Flattening rule:}\\

\begin{figure}[t]
\centering
\includegraphics[width=8cm,height=6cm,angle=0]{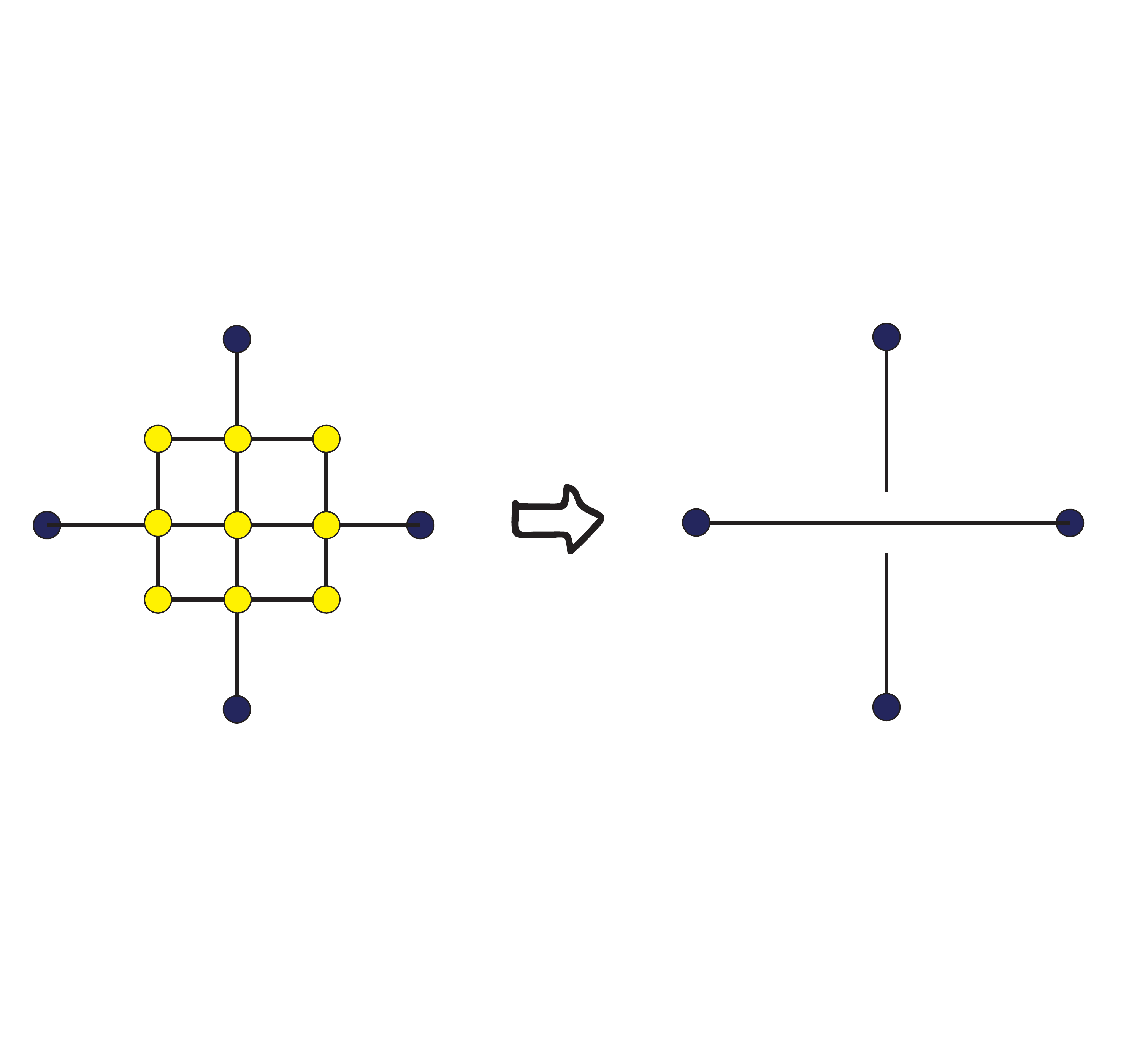}
\caption{(Color online) In the left hand, there is a graph state
where the yellow qubits are measured in the $Y$-basis,
successively. It leads to a new graph state with a crossing
between the edges on unmeasured qubits.} \label{knot2}
\end{figure}

It is well known that a single-qubit measurement on a qubit of a graph state in the $Y$-basis, leads to a new graph state that is local complement of the original one \cite{grstate}. By this fact, we give a rule that graphically shown in Figure (\ref{knot2}) where measurement on qubits which have been denoted by yellow circles in the $Y$-basis generates a new graph state with a crossing between two edges of the graph. Furthermore, the measurements should be performed step-by-step and it is necessary to apply an operator $\sqrt{Z}$ on the neighbors of the measured qubits in each step of the measurements. In order to correct probabilistic errors in each step, it is also enough to apply an operator $Z$ on the neighbors of the measured qubit.\\

Now we are ready to explain how the bipartite graph state can be generated by measurements on a 2D cluster state. A combination of uniformizing and flattening rules, face insertion and link insertion will convert the initial bipartite graph to a 2D rectangular lattice corresponding to 2D cluster state. Finally, we emphasize that the final 2D cluster state have several qubits more than the initial CSS topological state. In fact, our scheme involves five steps; cell insertion, flattening, uniformizing, face insertion and link insertion. It is clear that the number of added qubits in all the above steps grows polynomially with the number of qubits in the original topological model. In this way each topological CSS state can be efficiently generated by single-qubit measurements on a 2D cluster state. In the next section we test our scheme for two three-dimensional topological CSS states to show how it works.\\

\section{Examples }\label{ss}
By the measurement rules that we gave in the previous section, one will be able to follow a few graphical steps to find a measurement pattern corresponding to any CSS state. It is clear that the number of steps depends on the structure of the topological state and specially dimension of the underlying lattice. On the other hand, the CSS topological states in higher dimension are more important because of their application in universal topological quantum computation. Therefore, here we apply our scheme to two three-dimensional models namely the TC on a 3D rectangular lattice and the CC on a 3-colex. We also give some two-dimensional examples in appendix B that one can study before three-dimensional cases where the main idea can well be found.
\subsection{TC model on a 3D rectangular lattice}
\begin{figure}[t]
\centering
\includegraphics[width=10cm,height=7cm,angle=0]{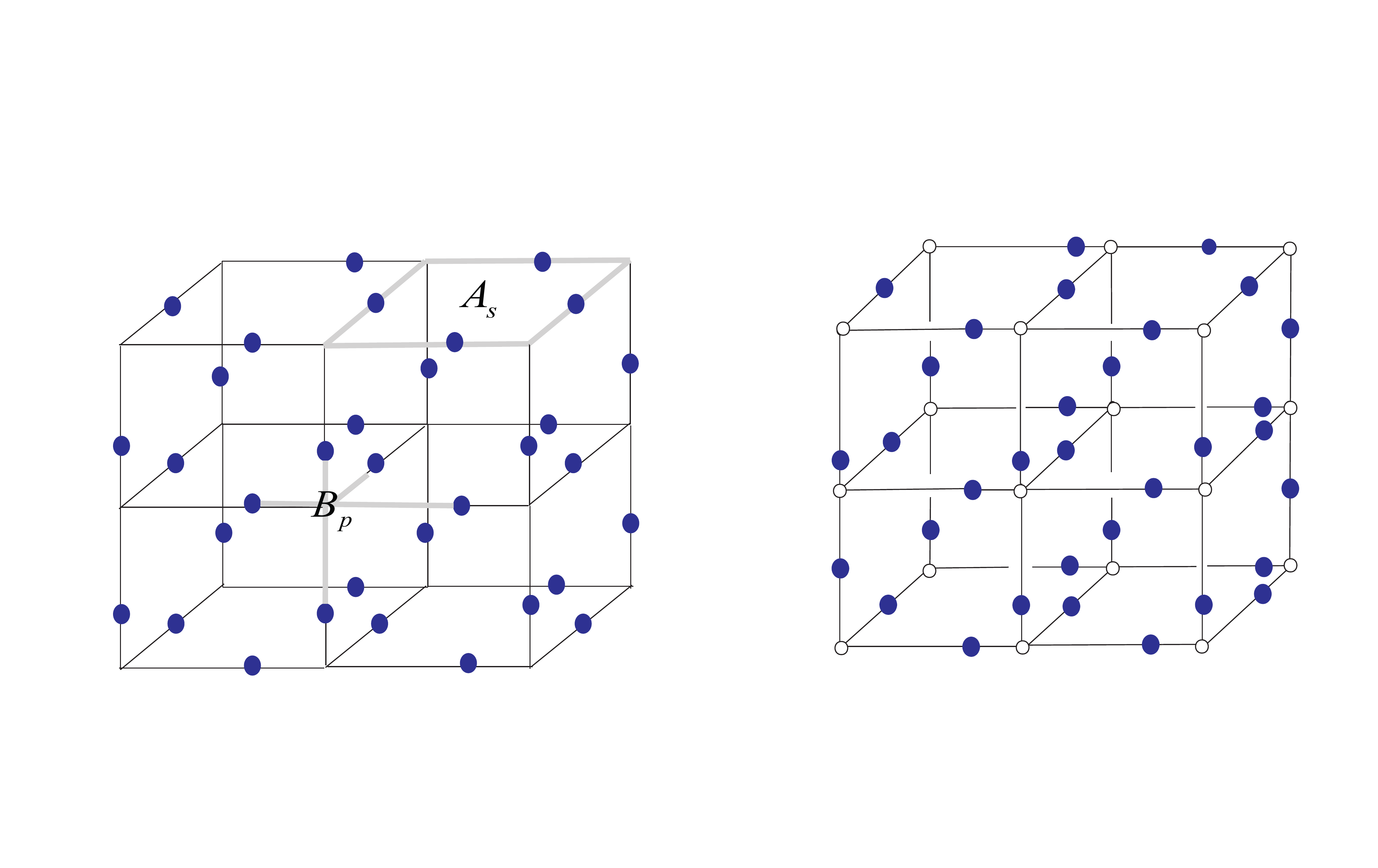}
\caption{(Color online), In the left-hand, we have a dual of TC on a 3D lattice where $X$-type operators are defined corresponding to plaquettes of the lattice and $Z$-type operators are defined corresponding to vertices of the lattice. In the right hand, we show a bipartite graph state corresponding to the CSS state in the left hand where white qubits should be measured in the $X$-basis.} \label{3Dtoric}
\end{figure}

In section (\ref{s1}) we introduced TC models on arbitrary graphs and specially on a 3D rectangular lattice, see Figure (\ref{3Dto}). Here we consider dual of that model where $Z$-type operators are defined corresponding to the vertices of the lattice and $X$-type operators are defined corresponding to the plaquettes of the lattice, see Figure (\ref{3Dtoric}, left). It is clear that this model can be converted to the original model by applying Hadamard operators $H$ on all qubits where $HXH=Z$ and $ZHZ=X$. We use the dual model because the process of correcting probabilistic errors is easier than the original model.\\

If we compare this model with the general structure in (\ref{q0}), it will be clear that each vertex of the 3D lattice corresponds to a $z$-cell. Therefore, in order to construct the corresponding bipartite graph state, it is enough to insert new qubits denoted by white circles in the vertices of the 3D lattice and connect them to black qubits living in adjacent edges. As it is shown in Figure (\ref{3Dtoric}, right), we will have a bipartite graph state on a 3D lattice where white qubits live in the vertices and black qubits live in the edges.\\

As we argued in the previous section, the result of measurements is probabilistic where it can lead to many excited cells. However, since we define the $z$-cells of the original model corresponding to vertices of the 3D lattice, there is a simple way to correct the probabilistic errors. In fact, each pair of the excited cells can be connect to each other by a string of $X$ operators which moves on edges of the lattice and we will be able to correct the errors by applying a string of $X$ operators between each pair of the excited cells.\\

\begin{figure}[t]
\centering
\includegraphics[width=10cm,height=7cm,angle=0]{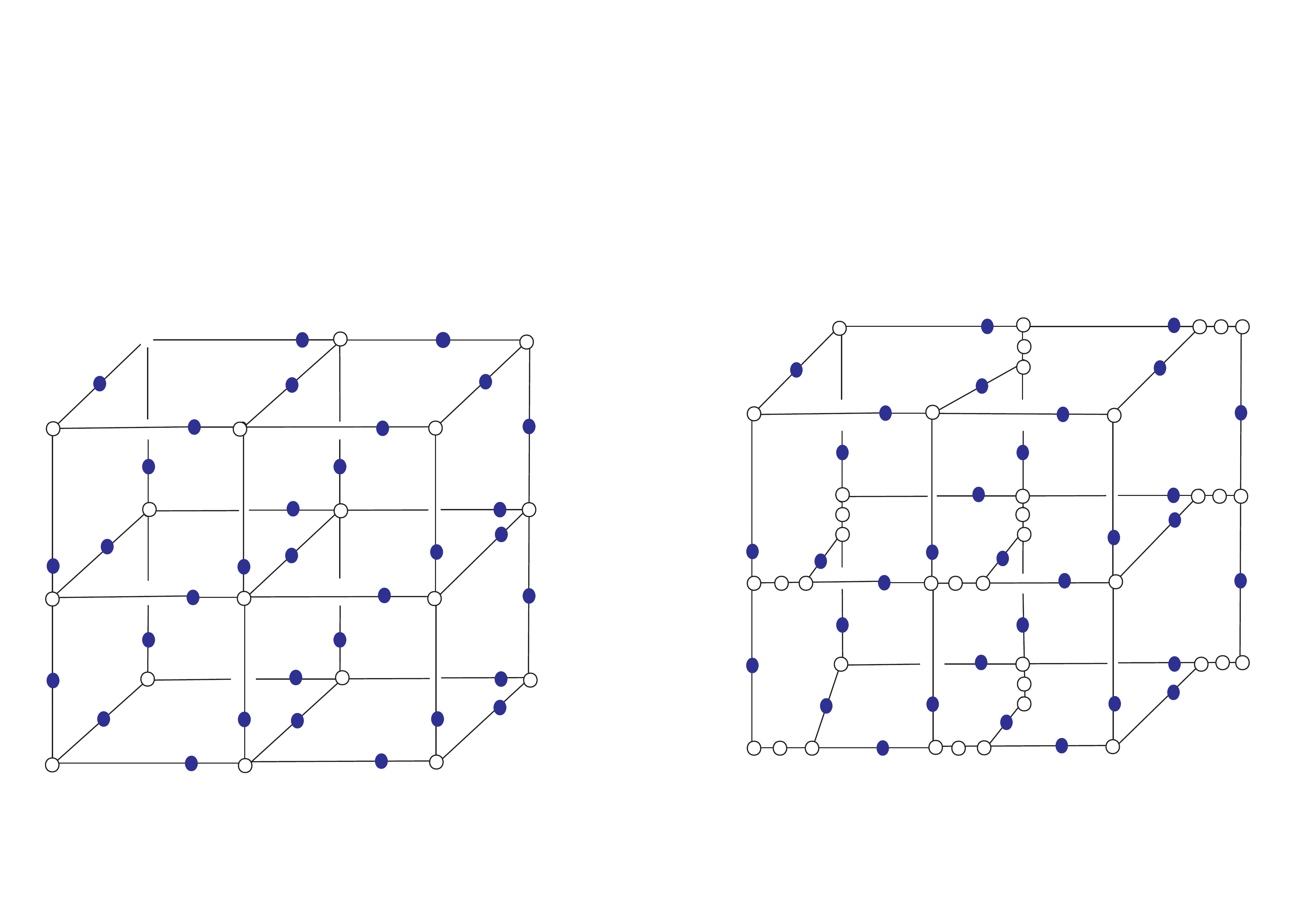}
\caption{(Color online) In the right hand, we use the uniformizing
rule to reduce degrees of all vertices in the left hand lattice.}
\label{cellinst}
\end{figure}
\begin{figure}[t]
\centering
\includegraphics[width=10cm,height=7cm,angle=0]{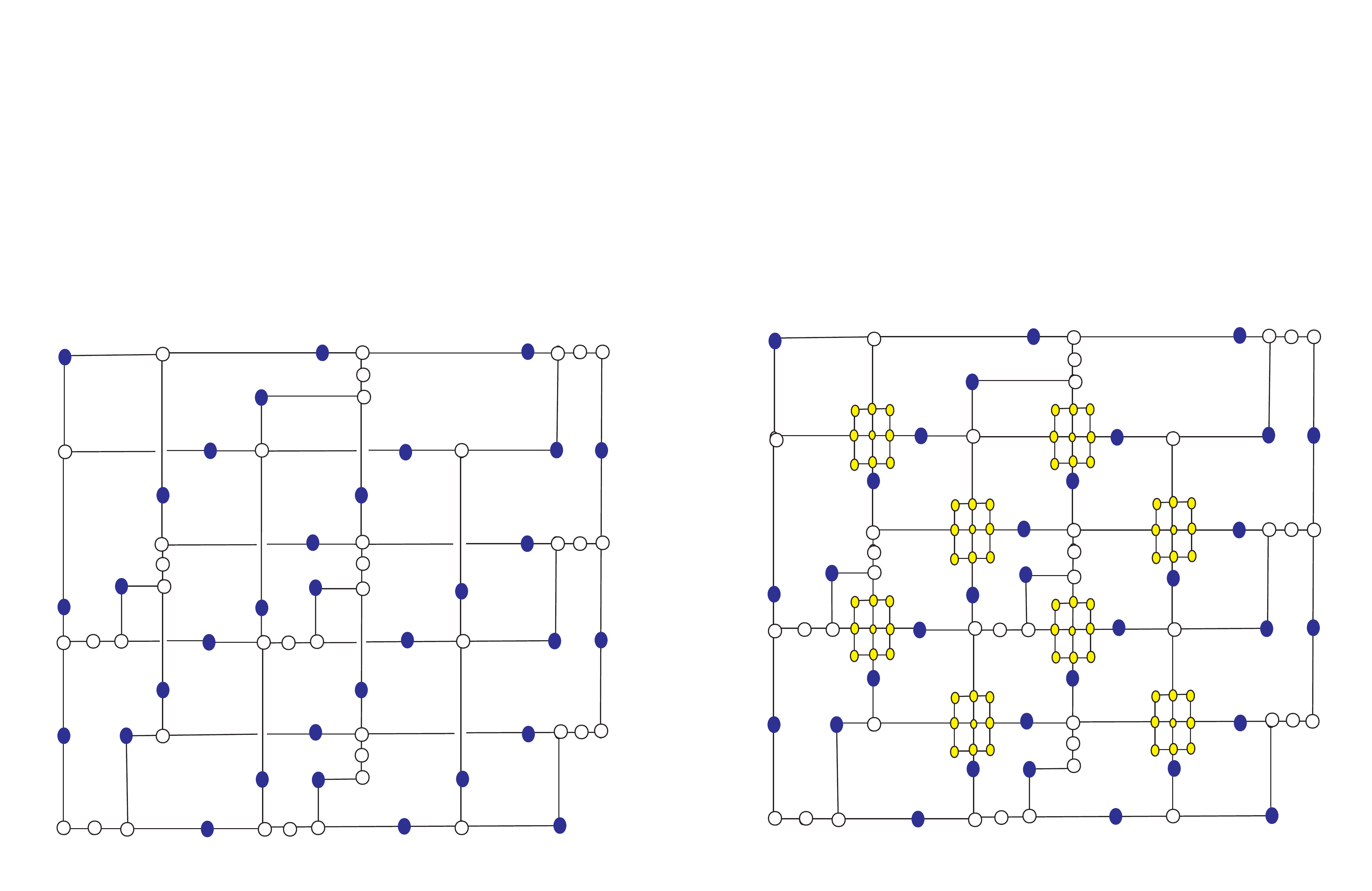}
\caption{(Color online) In the left hand, we deform all edges of
the lattice in (\ref{cellinst}-b) in order to convert the lattice
to a squarely structure. In the right hand, we use the flattening
rule to remove all crossings in the left hand lattice.}
\label{knott}
\end{figure}

So far we have found the corresponding bipartite graph state, see Figure (\ref{cellinst}-a). Now we use the flattening and uiformizing rules to convert it to a 2D cluster state. At the first, we apply the uniformizing rule as shown in Figure (\ref{cellinst}-b). Then by the flattening rule we can remove crossings between links, see Figure (\ref{knott}). Finally one can use the face insertion and the link insertion to fill all empty points to construct a 2D rectangular lattice.\\

\subsection{The CC on a 3-colex}
\begin{figure}[t]
\centering
\includegraphics[width=8cm,height=6cm,angle=0]{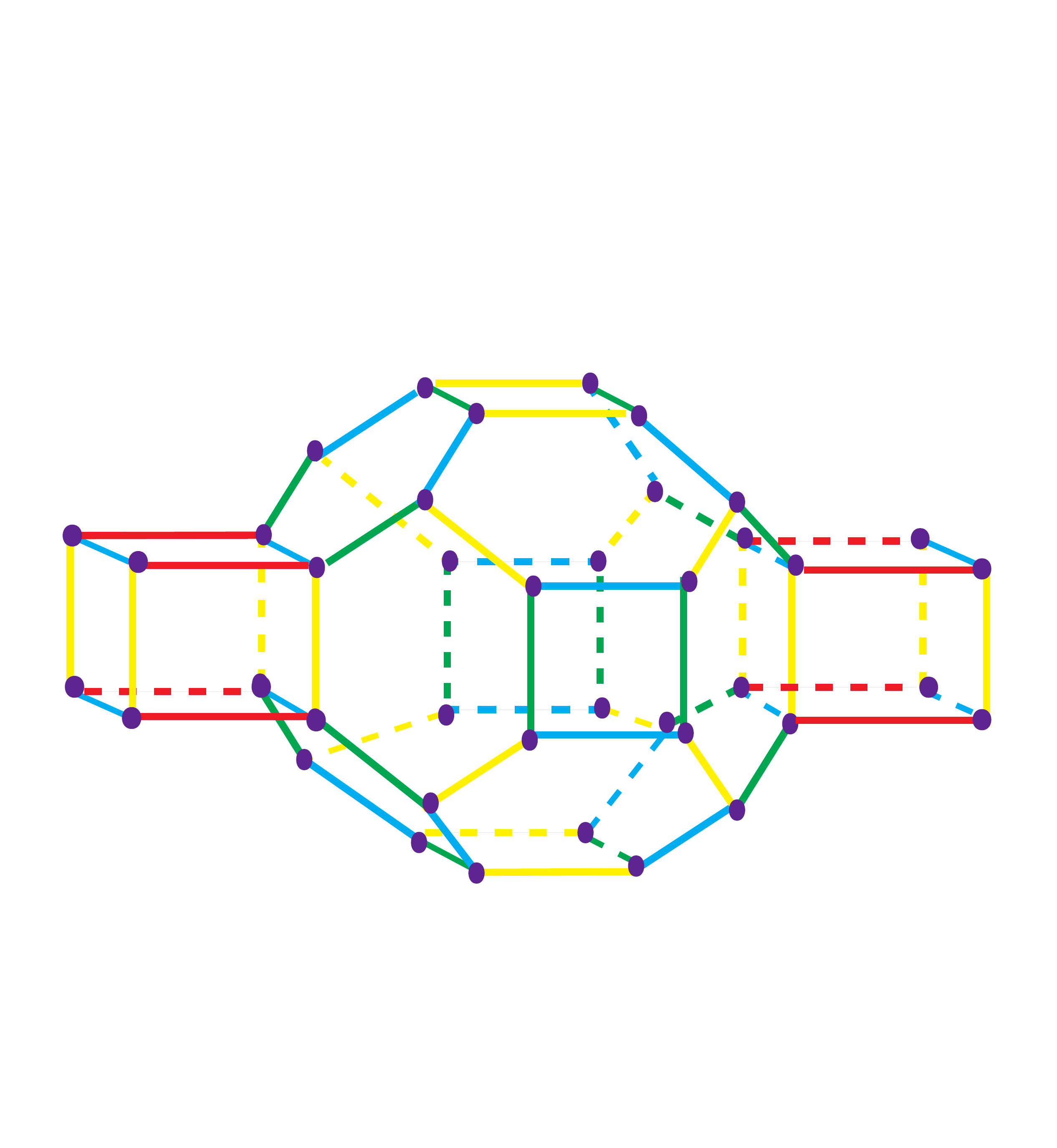}
\caption{(Color online) A 3-colex with three cells. The central cell involves 32 qubits and qubic cells involve 8 qubits.} \label{3Dexc}
\end{figure}
\begin{figure}[t]
\centering
\includegraphics[width=10cm,height=7cm,angle=0]{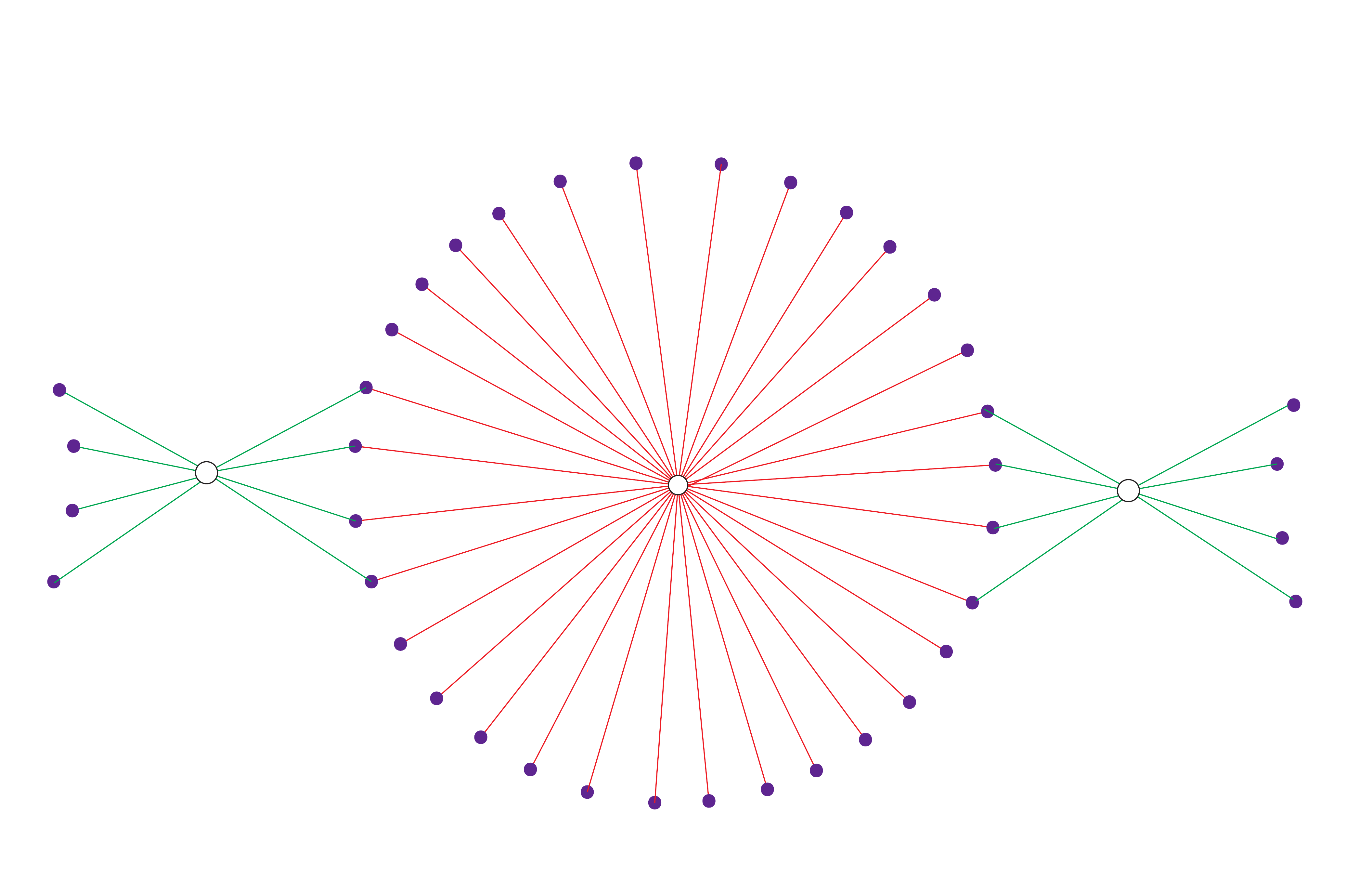}
\caption{(Color online) A bipartite graph state corresponding to CC model on 3-colex in Figure (\ref{3Dexc}). where white qubits should be measured in the $X$ basis.} \label{cellinsc}
\end{figure}

 Although the TC on a three-dimensional lattice is more important than two-dimensional case, it is not useful for a universal quantum computation. Therefore, here we consider a three-dimensional CC which has been shown that it can be served as a universal quantum computer. For simplicity, at the first we consider a 3-colex with only three cells as shown in Figure (\ref{3Dexc}) and then add other cells. Here, like TC on 3D lattice, we consider dual of the ordinary model where the $Z$-type operators are defined corresponding to the cells of the lattice and $X$-type operators are defined corresponding to the faces of the lattice. In this way, each cell of the 3-colex plays role of a z-cell in the general structure (\ref{q0}). In order to construct the corresponding bipartite graph state, we add a white qubit in the center of each cell and connect it to the black qubits of the corresponding cell, see Figure (\ref{cellinsc}).\\
 
Since we defined the $Z$-type operators corresponding to cells of the 3-colex, the process of correcting errors can be performed easily. In fact in a CC on 3-colex a string of $X$ operators on edges with a specific color generates excitations in cells with the same color in two endpoints of that string. Therefore, the errors generated in the excited cells of the lattice can be corrected by applying a string of $X$-operators between each pair of the excited cells with the same color.\\

\begin{figure}[t]
\centering
\includegraphics[width=12cm,height=8cm,angle=0]{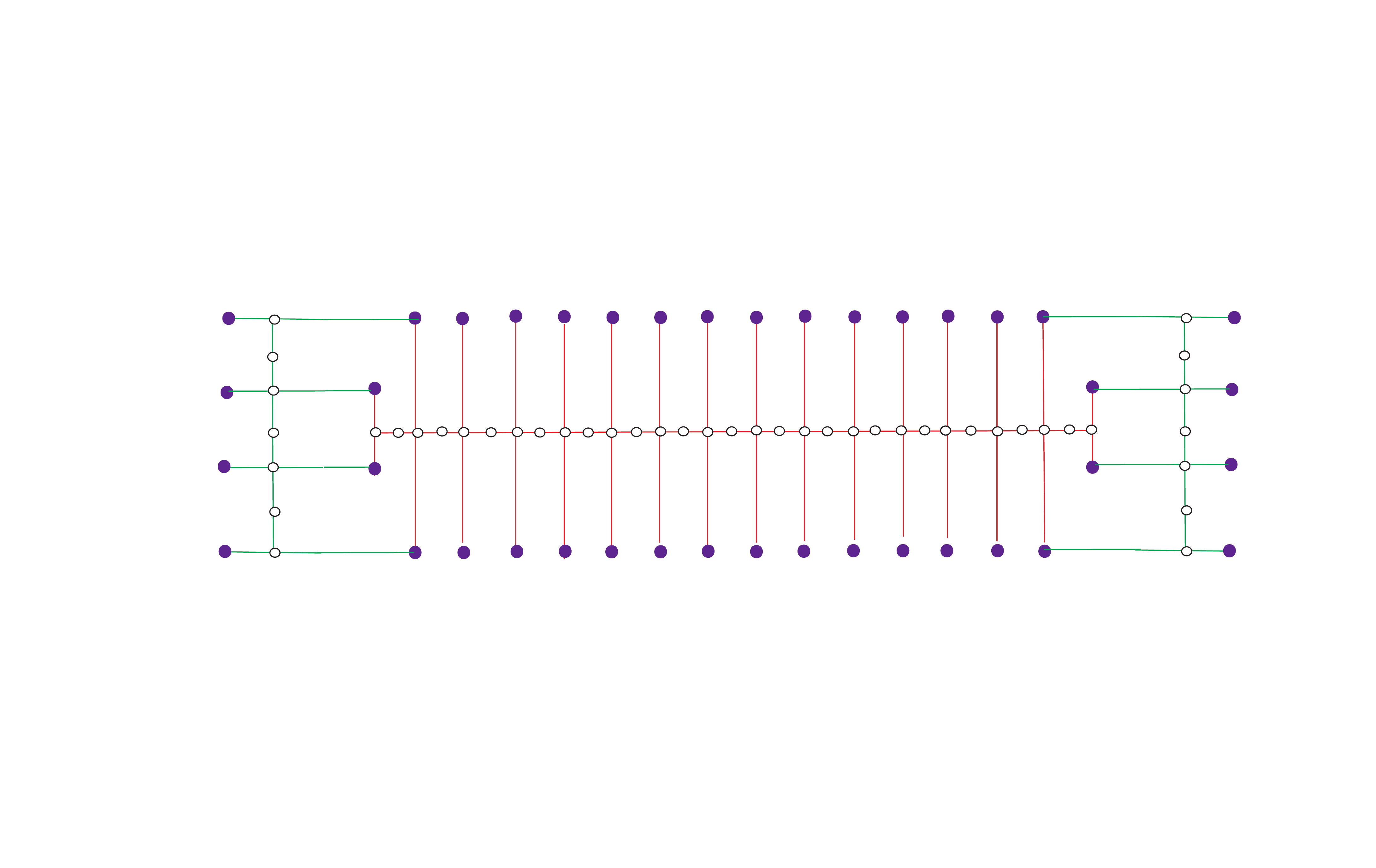}
\caption{(Color online) We use the uniformizing rule to reduce degrees of vertices. For the central cell in Figure (\ref{cellinsc}), we has used the uniformizing rule for sixteen times.} \label{mergcell}
\end{figure}
So far we have found the corresponding graph state to the CC on the 3-colex. As it is shown in Figure (\ref{cellinsc}), the degree of white vertices is very large. specially the central white vertex has 32 links. By the uniformizing rule, as it has been shown in Figure (\ref{mergcell}), we reduce degrees of the vertices to four. In this way the initial bipartite graph state converts to a 2D graph. Then by the face insertion and the link insertion, we can fill empty points to complete 2D rectangular lattice.\\

\begin{figure}[t]
\centering
\includegraphics[width=12cm,height=10cm,angle=0]{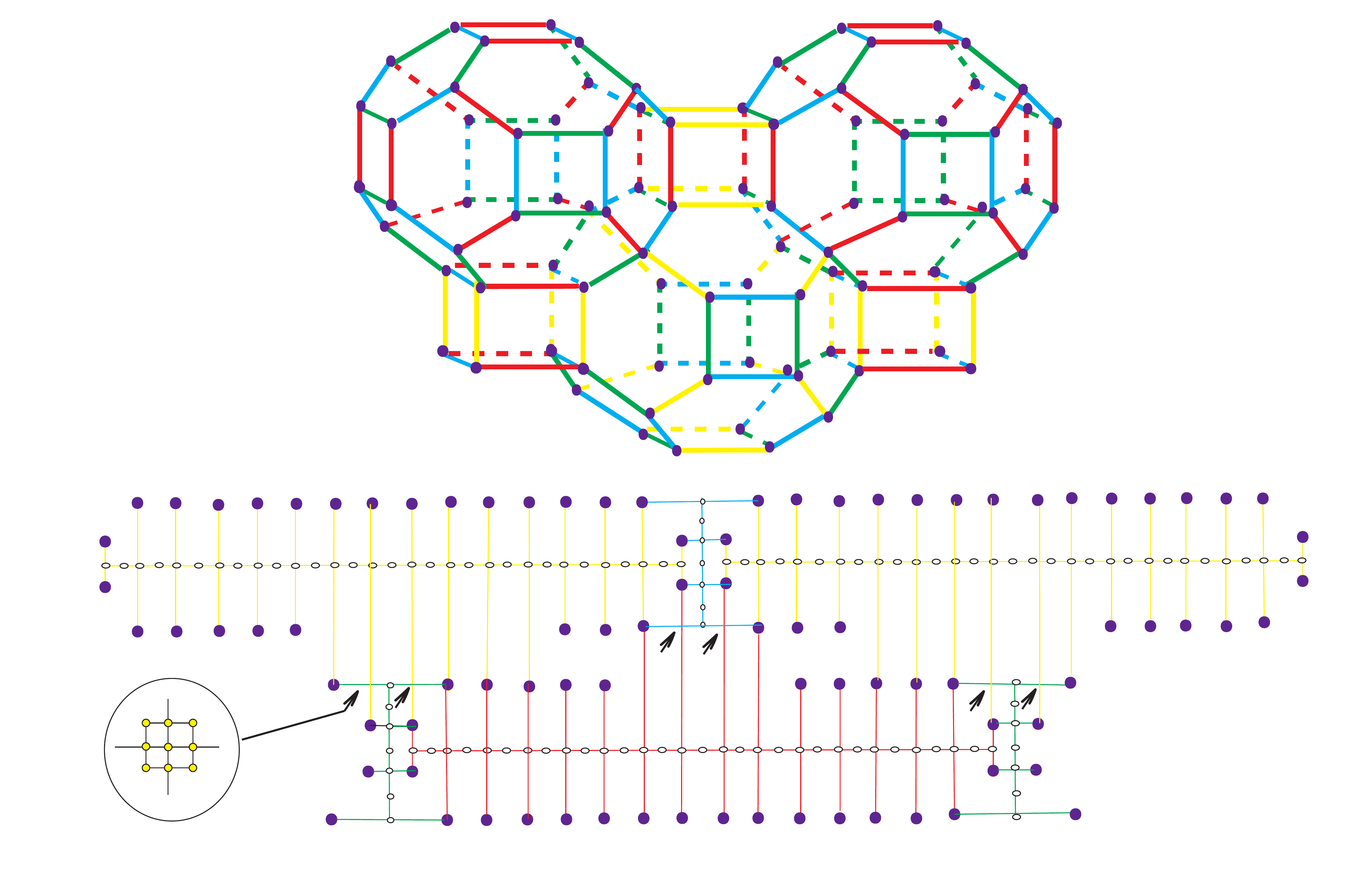}
\caption{(Color online) a) we show a 3-colex with six cells where dual of a CC has been defined. b) similar to the previous example in Figure (\ref{mergcell}), we use the uniformizing rule to reduce degrees of vertices. Furtheremore, there are also a few crossings between some edges that can be removed by the flattening rule.} \label{3Dexc2}
\end{figure}

 In Figure (\ref{3Dexc2}), we also repeat the above process for a bigger lattice with six cells. However, there are also some crossings between edges that we should use the flattening rule to remove them as shown in Figure (\ref{3Dexc2}). Finally, one can use the face insertion and the link insertion to complete the 2D rectangular lattice. It is clear that by adding other cells to the initial model, there will be more crossings that should be removed. However, the main idea is the same and our scheme can be applied for a CC on a 3-colex with more cells. \\
 
In this way, we gave a systematic way to convert each CSS topological state to a 2D cluster state by a proper measurement pattern. For topological states in higher dimension, the same method can be applied. However, the process of correcting probabilistic errors will be different where it depends on the structure of excitations of the model.
\section{Ising anyons as measurement-induced defects on the TC states}\label{s3}
In an interesting paper by Bombin, it has been shown that a dislocation in the lattice of a TC model on a square lattice leads to generate a pair of twists in two end-points of the dislocated line . He showed that when a charge anyon transports around a twist, it converts to a flux anyon. By such a property, Bombin showed that each twist plays the role of an Ising anyon. In this section we show that a single-qubit measurement on a TC on a 2D square lattice also leads to generate a pair of twists like what Bombin had shown. In this way, we propose a measurement pattern on cluster states to generate a topological state corresponding to the Ising anyons. \\

The Ising anyon model involves three particles which are named $1$, $\epsilon$ and $\sigma$ with the following fusion rules:
\begin{equation}
1\times \epsilon =\epsilon ~~,~~1\times \sigma =\sigma ~~,~~\epsilon \times \sigma =\sigma ~~,~~\sigma \times \sigma = 1 +\epsilon
\end{equation}
\begin{figure}[t]
\centering
\includegraphics[width=12cm,height=10cm,angle=0]{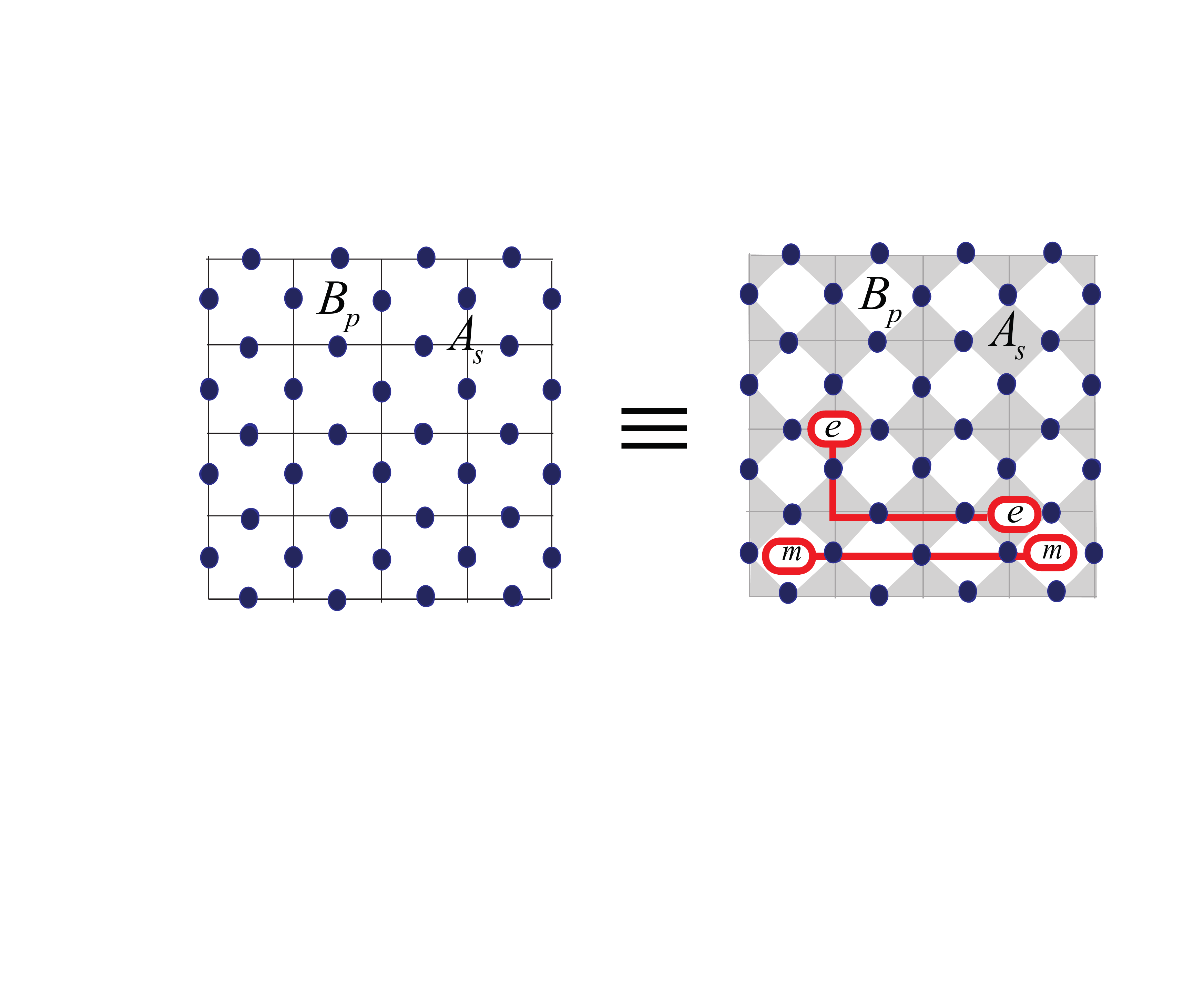}
\caption{(Color online), In the left-hand, there is a square lattice with qubits living on the edges of the lattice. In the right hand we have the same lattice where we attach a chess pattern on that lattice. A plaquette (vertex) operator $B_p$ in the left-hand corresponds to a dark (white) plaquette in the right-hand. A string operator in white (black) plaquettes generates two flux (charge) anyons in two endpoints of that string.} \label{kitaev}
\end{figure}
where $\epsilon$ is a fermion and $\sigma$ is called Ising anyon. \\

\begin{figure}[t]
\centering
\includegraphics[width=12cm,height=10cm,angle=0]{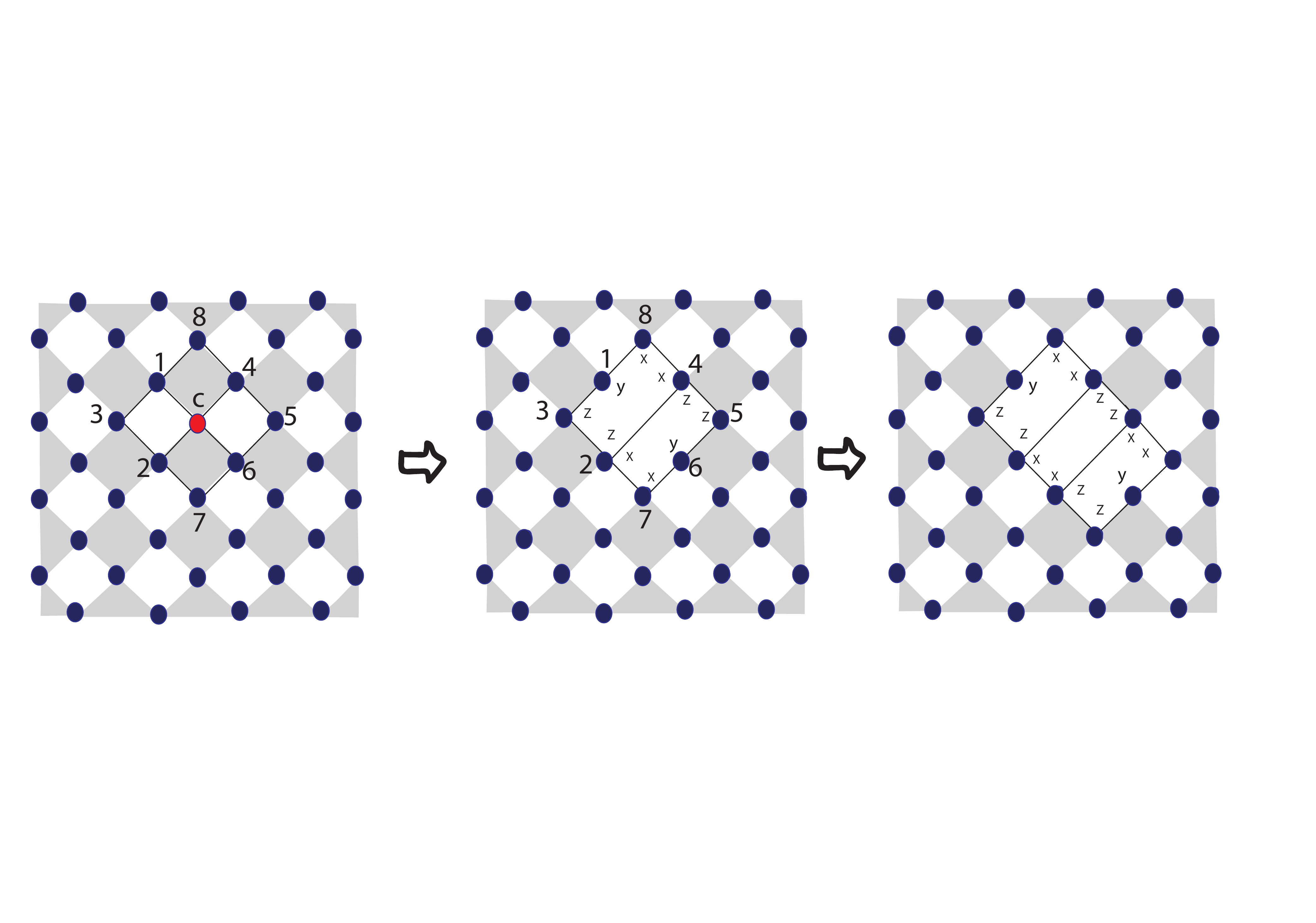}
\caption{(Color online) a) We perform a measurement on the qubit $a$ in the $Y$-basis. b) After measuring the qubit $a$, there are two new stabilizers corresponding to two deformed plaquettes. c) We transport one of the deformed plaquettes by measuring another qubit, that we have denoted by 6, in the $Y$-basis.} \label{ymeasure}
\end{figure}

Consider the toric code state on a square lattice. As it is shown in Figure (\ref{kitaev}), we can also show the stabilizers of the toric code state on a different square lattice where qubits live on the vertices instead of the edges. The plaquette operators correspond to the light plaquettes of the new lattice and the vertex operators correspond to the dark plaquettes of the new lattice. In the following, we measure one of the qubits, the qubit $c$ in Figure (\ref{ymeasure}-a), in the $Y$-basis. Since the Pauli operator $Y_c$ does not commute with four stabilizers of the TC state containing $B_p=Z_1 Z_c Z_2 Z_3$, $B_{p'}=Z_4 Z_5 Z_6 Z_c$, $A_s = X_1 X_8 X_4 X_c$ and $A_{s'}=X_c X_6 X_7 X_2$, these four operators should be removed from the set of stabilizers of the new state after measurement. But each product of them which commutes with the operator $Y_c$ can be a stabilizer of the new state. Therefore, the following operators are stabilizers of the new state
$$B_p B_{p'}=Z_1 Z_2 Z_3 Z_4 Z_5 Z_6~~,~~ A_s A_{s'}=X_1 X_8 X_4 X_6 X_7 X_2$$,
$$B_p A_s = (Z_2 Z_3 Y_1 X_8 X_4 )(-Y_c)~~,~~ B_{p'} A_s =(X_1 X_8 Y_4 Z_5 Z_6)(-Y_c)$$,
\begin{equation}\label{w1}
B_{p'} A_{s'}=(Z_4 Z_5 Y_6 X_7 X_2)(-Y_c)~~ and~~ B_p A_{s'}=(X_6 X_7 Y_2 Z_3 Z_1)(-Y_c)
\end{equation}
For exact determination of the new quantum state after measurement, it is necessary to find the generators of the stabilizers of that state. On the one hand, it is clear that only four numbers of the above operators are independent and we choose the first four of them to be:
\begin{equation}\label{sa}
B_p B_{p'}~~,~~ A_s A_{s'}~~,~~G:= B_p A_s = (Z_2 Z_3 Y_1 X_8 X_4 )~~,~~ \hat{G}: =B_{p'} A_{s'}=(Z_4 Z_5 Y_6 X_7 X_2)
\end{equation}
where the operator $Y_c$ which corresponds to the measured qubit $c$ has been suppressed. On the other hand, Since the toric code is defined on a torus with periodic boundary conditions, it is simple to show that the operator $B_p B_{p'}$ is equal to the product of other plaquette operators and also $A_s A_{s'}$ is equal to product of other vertex operators. In this way, among four operators in the relation (\ref{sa}), only the operators $G$ and $\hat{G}$ should be selected as the generators of the stabilizers of the new quantum state after measurement. We emphasize that although it was possible to choose two other operators $B_{p'} A_s$ and $B_p A_{s'}$, it leads to the same quantum state represented alternatively. Finally in a graphical notation we can show two new stabilizer operators $G$ and $\hat{G}$ by two deformed plaquettes in the lattice as shown in Figure (\ref{ymeasure}-b). Furethermore, we can also transport the deformed plaquettes by successive measurements in $Y$-basis. For example, in Figure (\ref{ymeasure}-c), we transport a deformed plaquette by measuring another qubit, denoted by 6 in the figure, in the $Y$ basis.\\

In the next step, we are ready to consider topological properties of the new quantum state after measurement. To this end, we should find the ground state of a new Hamiltonian as follows:
\begin{equation}\label{w2}
H=-\sum'_{p,s}(B_p +A_s ) -G -\hat{G}
\end{equation}
where $\sum'$ refer to all plaquettes of the lattice except of the two deformed ones in Figure (\ref{ymeasure}-b). Let us consider topological properties of this model by comparing it with the toric code state. The ground state of the toric code is a superposition of all loop constructions on the lattice. In fact, in a anyonic picture, the ground state of the toric code is a vacuum whose excitations are abelian anyons $1$, $e$, $m$, $\epsilon$, where any type of anyons can be created in pair, and each anyon can be moved on arbitrary path, anihilate the other anyon, and multiply the state by a phase.\\

 The excitations of the toric code model are also described by open strings. We define two kinds of strings living in the white and black plaquettes. each string can be represented by a binary vector whose components $a_j$ are equal to $1$ if string passes from a vertex called $j$ and are equal to zero otherwise. we denote such a binary vector by $a$. Corresponding to each string $a$ in the white plaquettes, we define an operator $S_x (a)=\prod_{i}X_{i}^{a_i}$. As it is shown in Figure (\ref{kitaev}), such a string of $X$ operators creates two charge anyons in two white plaquettes corresponding to two end-points of that string. Furthermore, corresponding to each string $a$ in the black plaquettes, we define an operator $S_z (a)=\prod_{i}Z_{i}^{a_i}$. Such a string of $Z$ operators creates two flux anyons in two black plaquettes related to two end-points of that string as shown in Figure (\ref{kitaev}).\\
 
\begin{figure}[t]
\center
\includegraphics[width=10cm,height=10cm,angle=0]{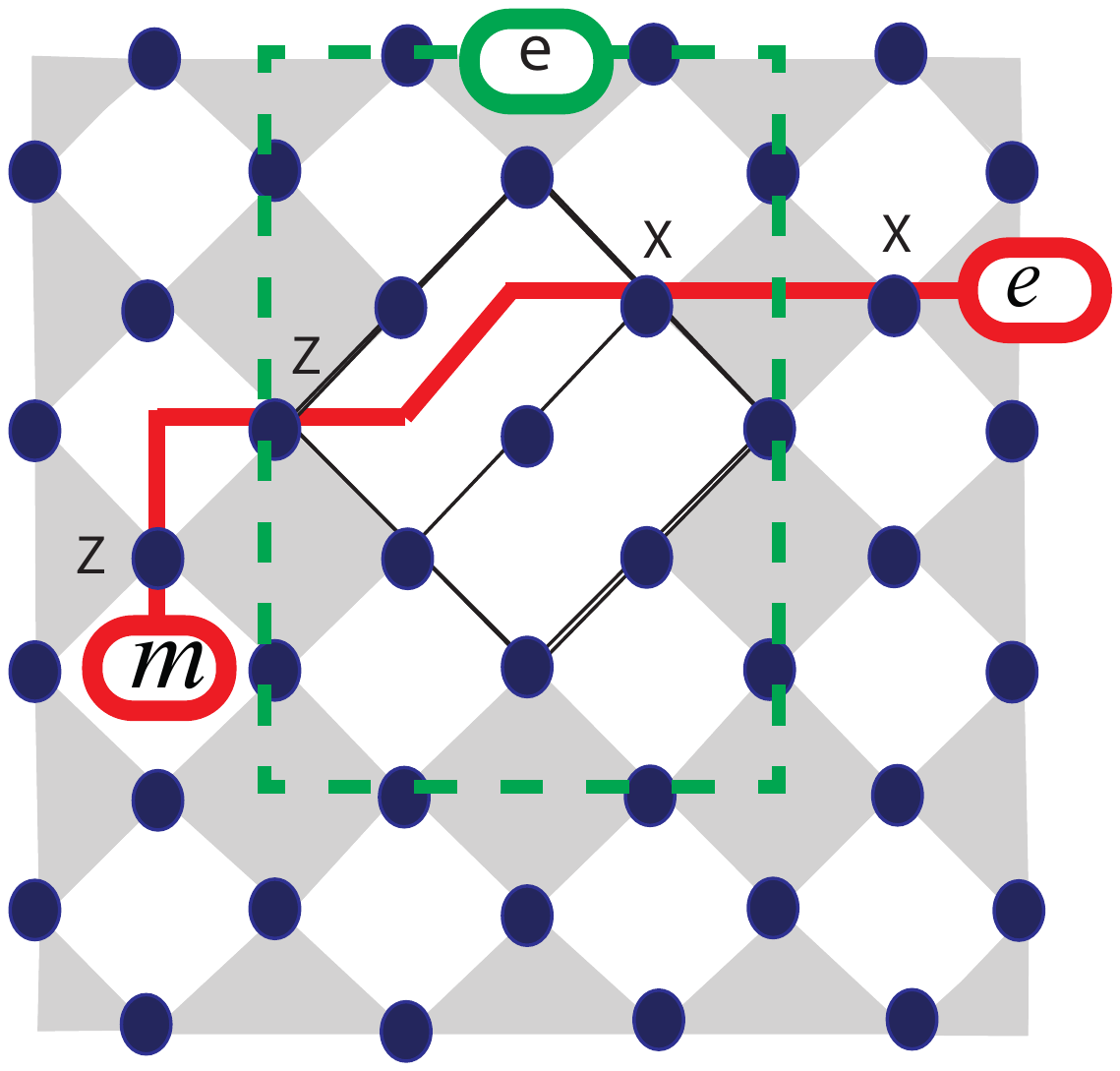}
\caption{(Color online) Red curve; a flux particle is converted to
a charge particle when transport around a twist. Green loop; if a charge
particle turns around both of twists, it can generate a closed
string. } \label{twist}
\end{figure}
The open strings in the toric code model had an important property that can be labeled by two different colors. This property is completely related to anyonic picture in the toric code model where there are two different anyons labeled by charge and flux anyons. However, the situation is different for the new model in (\ref{w2}) where there are also other open strings with different property. As it has been shown in Figure (\ref{ymeasure}), measurement in the $Y$ basis generates a pair of deformed plaquettes that can be transported by successive measurements. Such an operation in the lattice can be denoted by a line that connects the two deformed plaquettes and shows the trajectory of $Y$ measurements, we call it a cutting line. By this fact, as shown in Figure (\ref{twist}), we can also define a new open string operator that crosses the cutting line where it is generated by applying $X$ and $Z$ operators as shown in the figure. This open string has two end-points in a white and a black plaquette corresponding to a charge and a flux anyon in the toric code model. It means that in the new model we can not label anyons by two different colors. In fact, a charge anyon is converted to a flux anyon when it is transported around one of the deformed plaquettes. According to definition of Bombin in \cite{9}, such a property of the deformed plaquettes shows that each one of them plays role of a twist for anyons of the toric code model. \\

\begin{figure}[t]
\center
\includegraphics[width=10cm,height=8cm,angle=0]{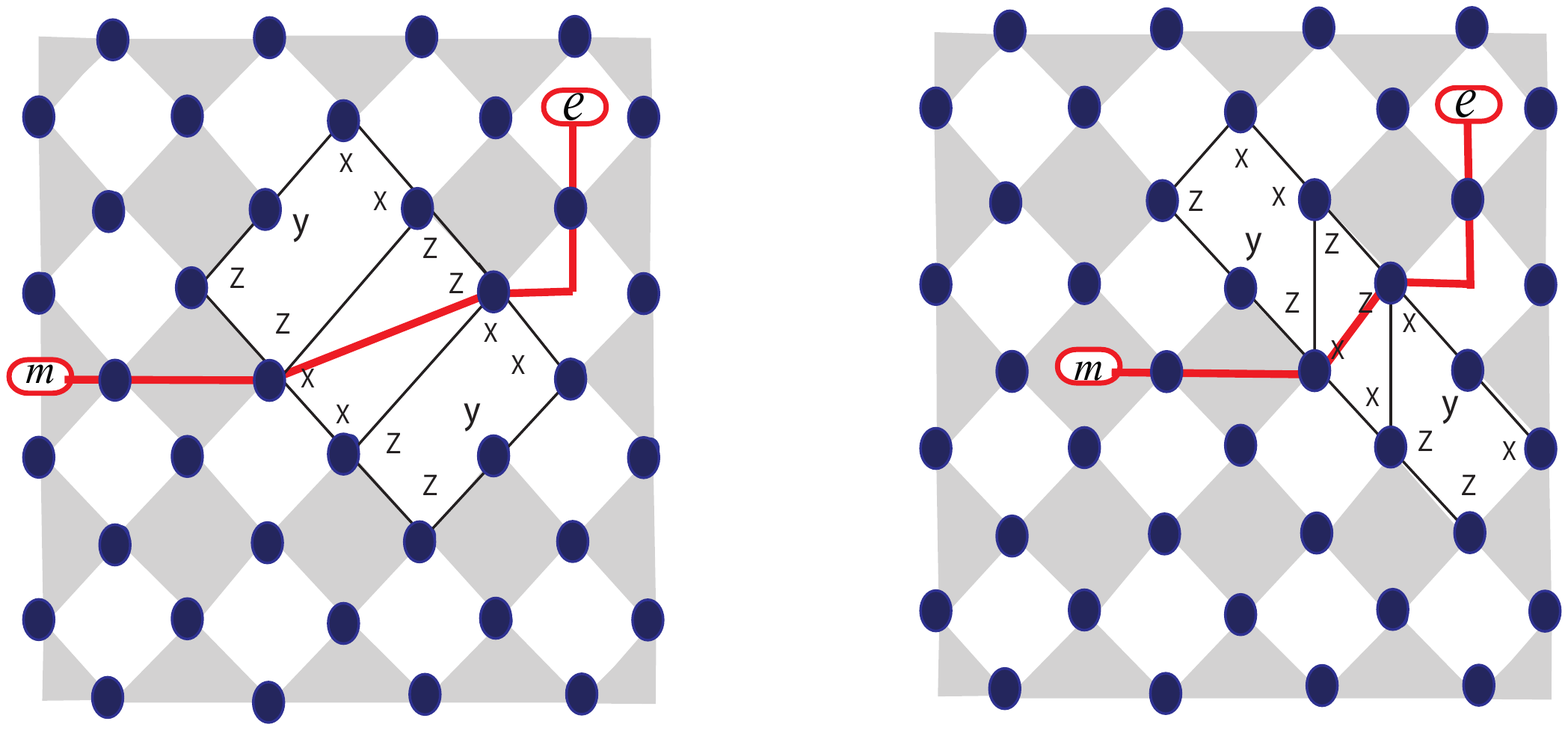}
\caption{(Color online) In the right hand, we have shown a pair of twists which have been generated by dislocation in the lattice and have been proposed by Bombin at \cite{9}. In the left hand, we have shown a pair of twists which have been generated by measurement. In both models a charge anyon converts to a flux anyon when it transports around a twist. } \label{tornew}
\end{figure}
In this way, measurement-induced defects that are generated by
measurement in $Y$- basis on the toric code state play role of the
same twists that Bombin had already considered by dislocations in
the lattice, see Figure (\ref{tornew}). According to the Bombin's
work, such twists are equal to the Ising anyons which are
non-abelian anyons. Although one can find a complete argument in
\cite{9}, we briefly explain, in appendix C, how the
fusion rule of Ising anyons hold for the twists. We emphasize that
although twists in our model have the same properties of the
twists in \cite{9}, our model has a different approach to generate
the twists where it is possible to realize them by single-qubit
measurements on a cluster state.
\section{Discussion}
In this paper we considered three sets of important topological states containing TC states on different topological structures, CC models on different lattice in various dimensions and Ising anyons which were realized as topological defects on the TC states. We gave a general scheme for preparation of the above states by single-qubit measurement on cluster states. By a set of graphical transformations, we showed how one can find the measurement pattern on the cluster states corresponding to different topological states. Specifically, we found a specific measurement pattern for realization of Ising anyons.
\section*{Acknowledgement}
I would like to specially thank V. Karimipour for his kindly helps and good comments during preparation of the paper.

\section{Appendix A}\label{A1}
In this appendix, we give an explicit proof that shows how a CSS
quantum state can be generated by single-qubit measurements in the
$X$-basis on a bipartite graph state. To this end consider a
general CSS state in the following form:
\begin{equation}\label{qa}
|\psi \ra =\prod_{c}(1+C_z)|++...+\rangle
\end{equation}
where $c$ denotes a $z$-cell of the model. We use the fact that $C_z =\prod_{i\in c} Z_i$ and prove the following lemma to rewrite the operator $1+C_z $:\\

\textbf{Lemma}: \\

\begin{equation}\label{x5}
1+C_z =\sum_{s_c = -1,1} e^{i\pi (\frac{1-s_{c}}{2})(\sum_{i\in c} \frac{1-Z_{i}}{2})}
\end{equation}
\textbf{proof:} For the left hand of the relation, By the fact that $Z_i = e^{i\pi \frac{1-Z_{i}}{2}}$, we can write $C_z  =e^{i\pi \sum_{i\in c} \frac{1-Z_{i}}{2}}$. Finally, it is enough to span the summation $\sum_{s_c =-1,1}$ in the right hand of the relation to confirm the Lemma.\\

If we replace the relation (\ref{x5}) in the CSS state (\ref{qa}), we will have:
\begin{equation}\label{x6}
|\psi\ra =\prod_{c} \{\sum_{s_c = -1,1} e^{i\pi (\frac{1-s_c}{2})(\sum_{i\in c} \frac{1-Z_{i}}{2})}\} |+++...+\rangle
\end{equation}
where corresponding to each $z$-cell $c$ we have defined a new variable $s_c$. In the next step, we use a simple equation that for a function of variable $S=\{-1,1\}$ as $f(s)$. We have $\sum_{s=-1,1}f(s)=2 \la+|f(Z)|+\ra$, where $Z$ is the Pauli operator. If we use this fact in the relation (\ref{x6}), we will have a new form of the CSS state in the following form:
\begin{equation}\label{q3}
|\psi \ra =2^{N_c}\prod_{c} \{ \la +|_c e^{i\pi (\frac{1-Z_c }{2})(\sum_{i\in c} \frac{1-Z_{i}}{2})}|+\ra _c \} |+++...+\rangle
\end{equation}
where $N_c $ is the number of $z$-cells and $|+\ra_c$ refers to a new qubit which has been added on the center of each $z$-cell. In above relation, let us consider the operator $ e^{i\pi (\frac{1-Z_c }{2})(\sum_{i\in c} \frac{1-Z_{i}}{2})}$. It is clear that this operator can be also written as $\prod_{i\in c} e^{i\pi (\frac{1-Z_c }{2})(\frac{1-Z_{i}}{2})}$. Furthermore, it is simple to show the operator $ e^{i\pi (\frac{1-Z_c }{2})(\frac{1-Z_{i}}{2})}$ is equal to the well-known operator $CZ_{c,i}=|0\ra_c \la0|\otimes I +|1\ra_c \la 1| \otimes Z_i$, therefore we will have:
\begin{equation}\label{q4}
|\psi\ra =2^{N_c}\prod_{c}\prod_{i \in c} \{\la+|_p CZ_{c,i}|+\ra_c\} |+++...+\rangle
\end{equation}

By above process, we convert the CSS state in the relation (\ref{qa}) to the inner product of two states in the form of the the above relation. We can give a graphical notation to such a transformation. To this end, we denote qubits of the initial model by black circles, see Figure (\ref{sexam}) for a specific example. In the first step, corresponding to the relation (\ref{x6}) we insert a new qubit $S_c$ corresponding to each $z$-cell, we denote these qubits by white circles. Then we connect each white qubit $S_c$ to other black qubits belonging to the $z$-cell $c$. In this way, we will have a bipartite graph where black qubits of each $z$-cell of the initial model are connected to a white qubit corresponding to the same $z$-cell, see Figure (\ref{sexam}).\\

In the following, let us return to the relation (\ref{q4}) and interpret it by the above graphical notation. In fact, operator $CZ_{c,i}$ have been applied between a white qubit and a black qubit corresponding to each edge of the bipartite graph. Since operators $CZ_{c,i}$ have been applied to a product state as $|++...+\ra$, the state of $\prod_{c}\prod_{i \in c} CZ_{c,i}|+\ra_c |+++...+\rangle $ is a graph state which is defined on the bipartite graph. Therefore, we can consider the CSS state $|\psi\ra$ in (\ref{q4}) as a product of the state $\prod_{c} \la+|_c$ and a graph state where all white qubits of the graph state are projected on state $|+\ra$. This is equivalent to measuring white qubits in the $X$-basis when the positive eigenvalue is derived.\\

\begin{figure}[t]
\centering
\includegraphics[width=12cm,height=8cm,angle=0]{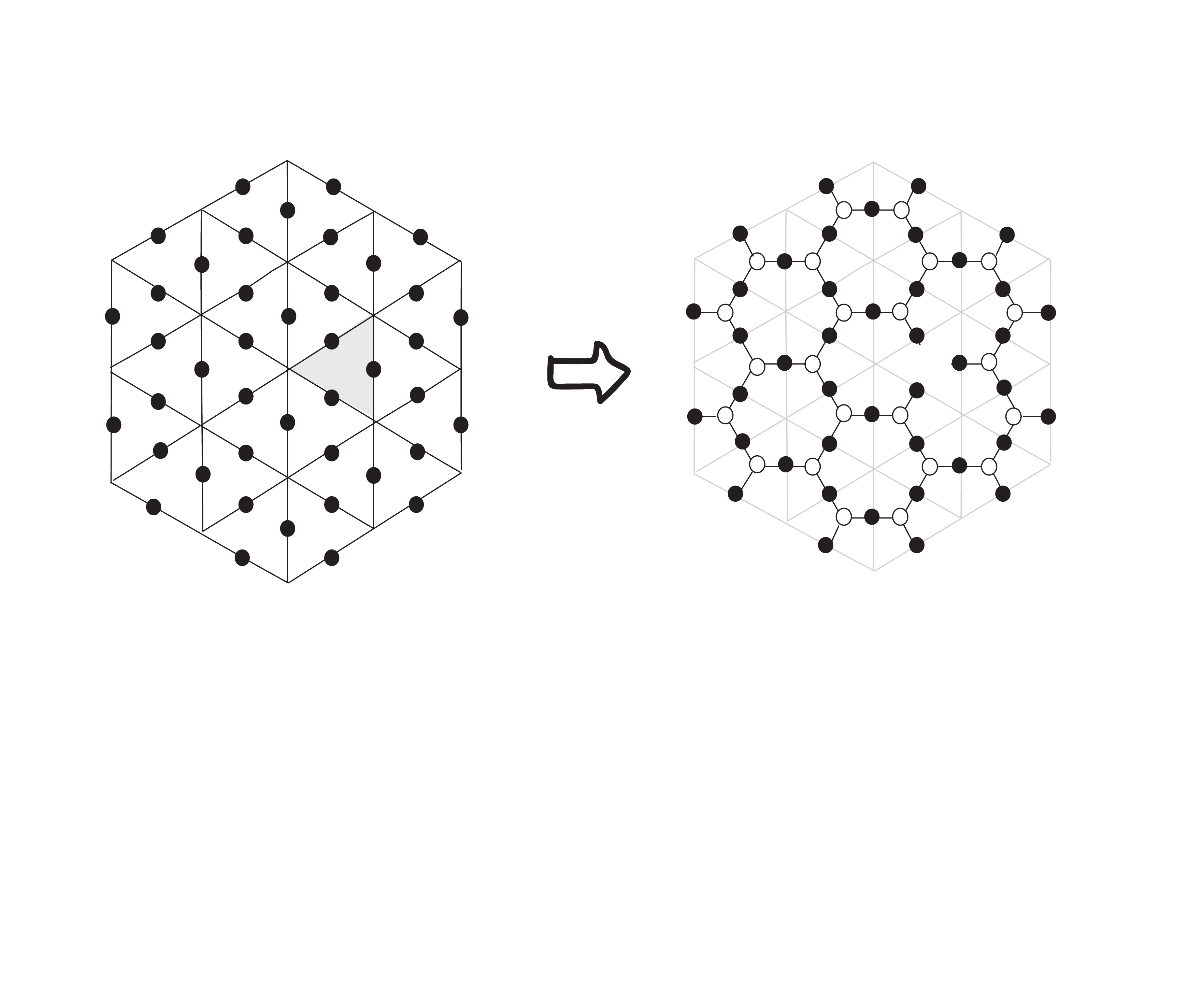}
\caption{(Color online) In the left hand, there is a TC state on a triangular lattice. the gray triangle show a hole in the lattice where the corresponding plaquette operator is absent. In the right hand, there is a graph state on a bipartite graph where white qubits should be measured in the basis of the Pauli operator $X$.} \label{ex1-k}
\end{figure}
It is useful to show how the above approach works for the TC states and the CC states on some simple graphs. To this end, we consider a TC state on a two-dimensional triangular lattice with a hole in the center of lattice and also we consider a CC state on a hexagonal lattice.\\

As the first example, in Figure (\ref{ex1-k}) we show a triangular graph where qubits live in the edges of the graph, we denote these qubits by black circles. The TC state on such a graph is defined by defining plaquette and vertex operators corresponding to each plaquette and vertex of the graph. We consider a hole in the lattice by removing a plaquette operator corresponding to a plaquette of the lattice that is denoted by gray color in Figure (\ref{ex1-k}). According to the lemma (\ref{x5}), we add new qubits in the center of all plaquettes of the lattice except of the gray plaquette, we denote new qubits by white circles. Then, we connect each new qubit to three qubits that live in the edges of the corresponding plaquette, Figure (\ref{ex1-k}). If we remove the initial triangular lattice, we will have a new graph as it is shown in Figure (\ref{ex1-k}). According to the relation (\ref{q4}), we define a graph state corresponding to this new graph and measure all white qubits in the $X$-basis. In this way, we have the initial TC state on black qubits when the positive eigenvalue of the operator $X$ is derived.\\

\begin{figure}[t]
\centering
\includegraphics[width=10cm,height=7cm,angle=0]{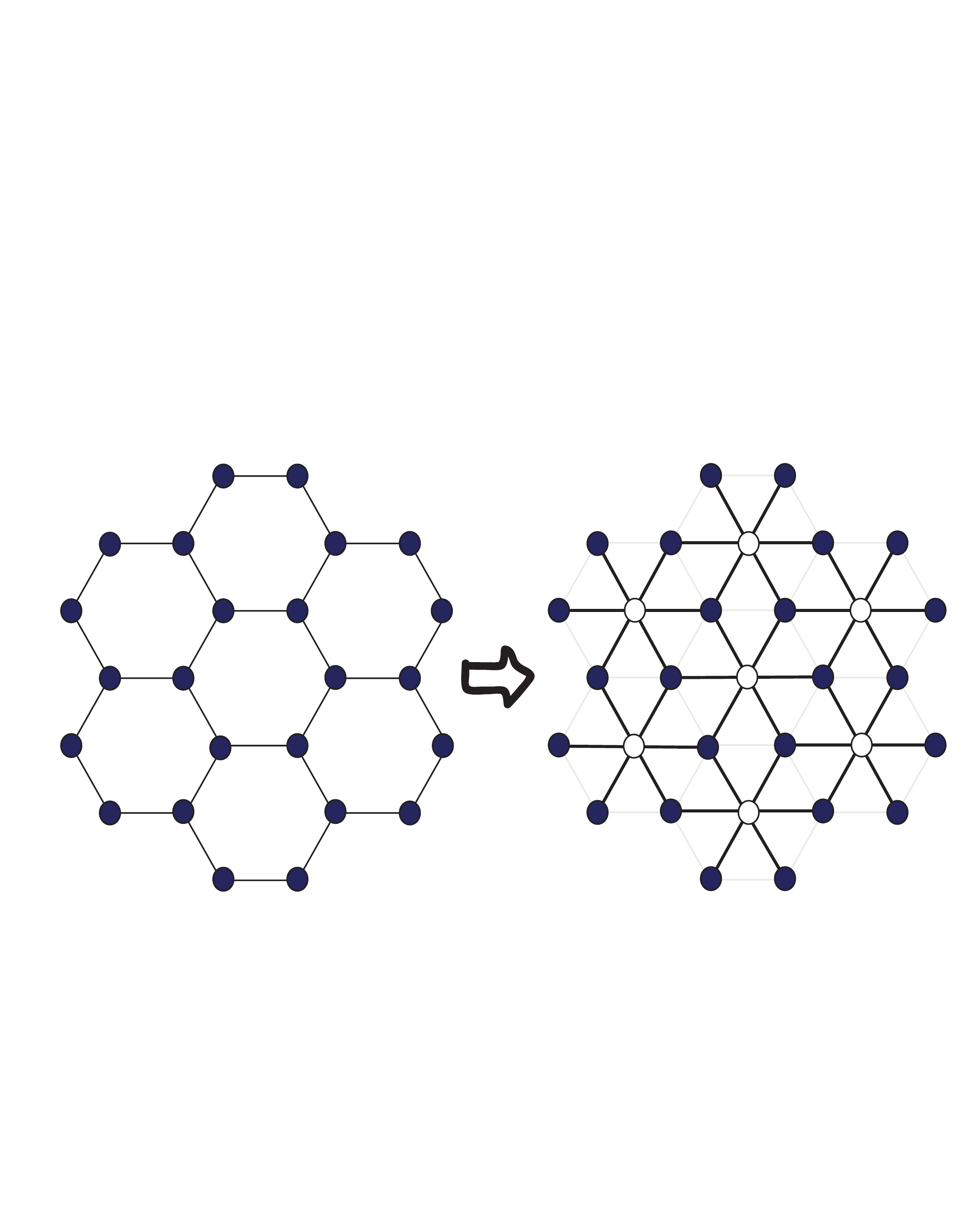}
\caption{(Color online) In the left hand, there is a CC state on a
hexagonal lattice. In the right hand, there is a graph state on a
bipartite graph where white qubits should be measured in the basis
of the pauli operator $X$. } \label{ex1-c}
\end{figure}
As the second example, in Figure (\ref{ex1-c}), we consider a
hexagonal lattice where qubits live in the vertices of the graph
and we denote them by black circles. The CC state is defined by
considering the plaquette operators $h_x$ and $h_z$ corresponding
to all plaquettes of the lattice. According to the lemma
(\ref{x5}) we add new qubits denoted by white circles in the
center of all plaquettes of the lattice. Then, as it is shown in
Figure (\ref{ex1-c}), we connect each white qubit to all six black
qubits that live in the vertices of the corresponding plaquette.
In this way, after removing the initial hexagonal lattice, we have
a bipartite graph with white and black qubits. According to the relation
(\ref{q4}), we define a graph state on such a graph. If we measure
all white qubits in the $X$-basis, we will have the initial CC
state on white qubits when the positive eigenvalue of the operator
$X$ is derived.
\section{Appendix B}\label{B1}
In this Appendix, we give two simple two-dimensional examples of the CSS topological states and show how they can be generated by single-qubit measurements on a cluster state. To this end, we consider the same examples given in appendix A. In Figure (\ref{ex2-k}), we show the corresponding graph state to the TC state on triangular lattice. Since the degree of all vertices in the graph are lesser than four, we can easily match the graph on a square lattice, see Figure (\ref{ex2-k}). In order to fill the empty points on the square lattice, we use the face insertion to add qubits on the empty points. We show these qubits in Figure (\ref{ex2-k}) by blue circles. In this way, we have a cluster state that white circles has been measured in the $X$-basis and the blue circles are measured in the $Z$-basis. The final state on unmeasured qubits will be the initial TC state.\\

\begin{figure}[t]
\centering
\includegraphics[width=12cm,height=8cm,angle=0]{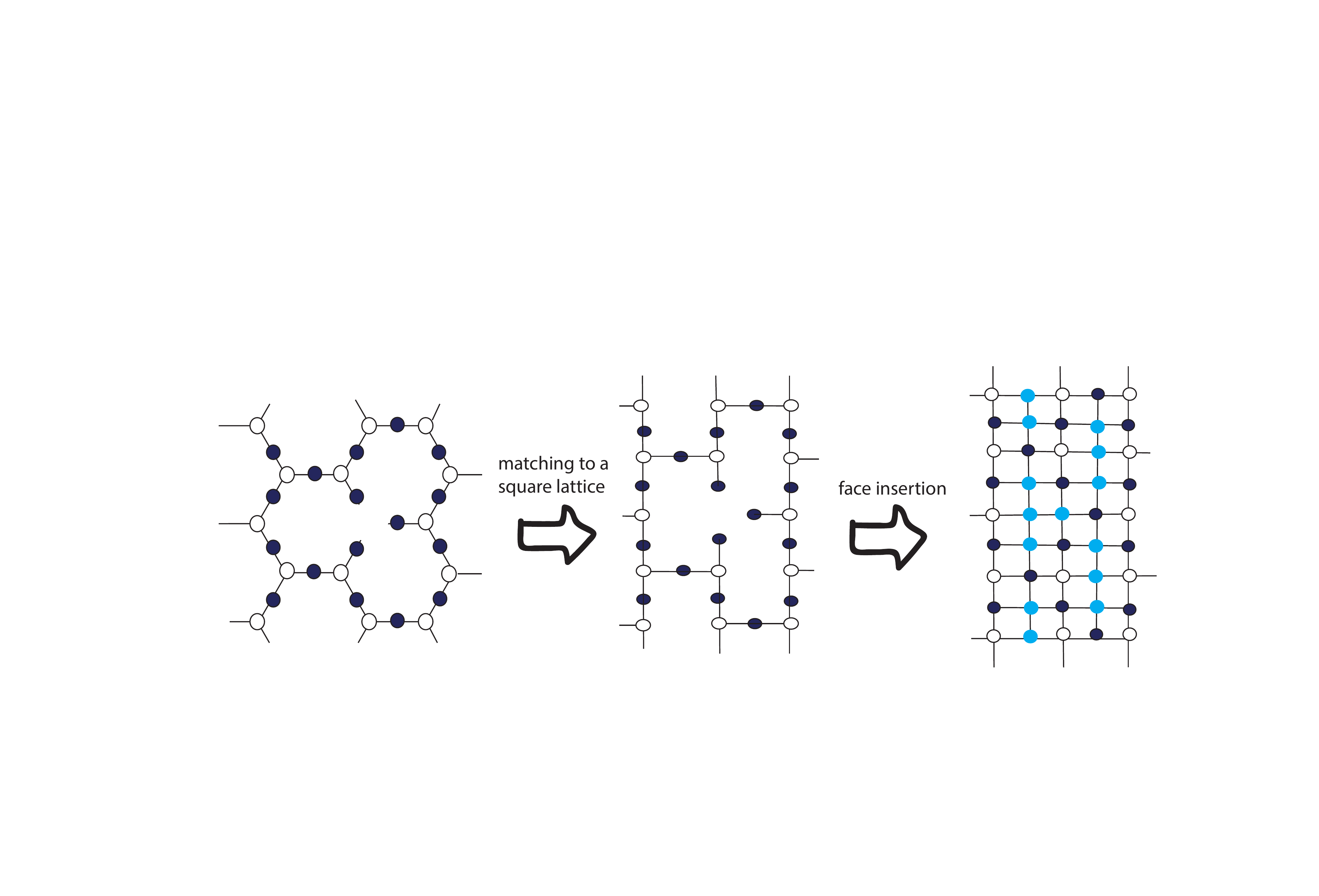}
\caption{(Color online) In the left hand, there is a graph state on a bipartite graph corresponding to the TC state on a triangular lattice. where the white qubits should be measured in the basis $X$. In the center, by deforming the edges, the initial graph is matched on a square lattice where the white qubits should be measured in basis $X$. In the right hand, By the face insertion the empty points are filled by blue qubits that should be measured in the $Z$-basis. } \label{ex2-k}
\end{figure}
\begin{figure}[t]
\centering
\includegraphics[width=12cm,height=8cm,angle=0]{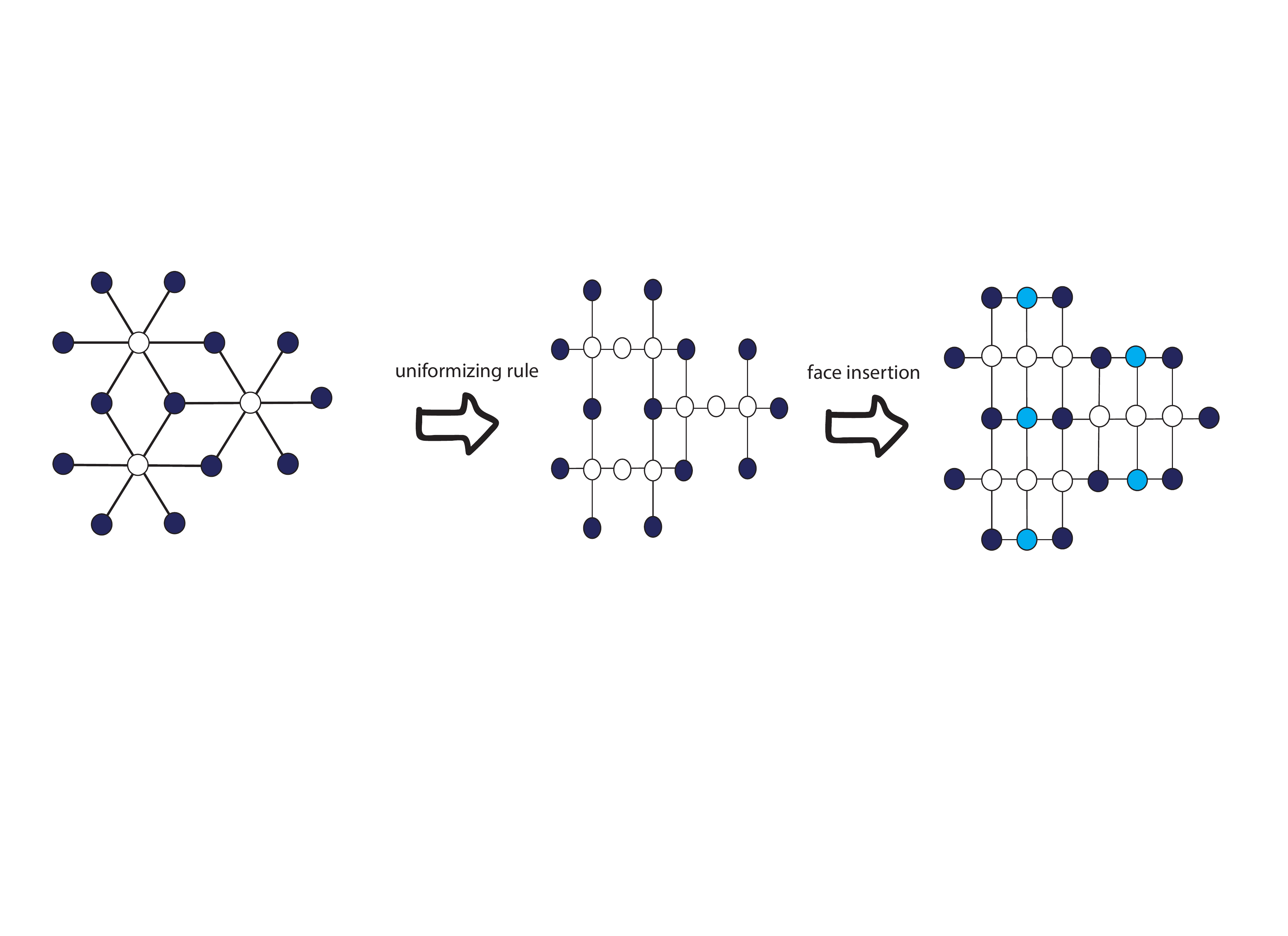}
\caption{(Color online) In the left hand, there is a graph state on a bipartite graph corresponding to the CC state on a hexagonal lattice. where the white qubits should be measured in the basis $X$. In the center, by merging rule, the initial graph is matched on a square lattice where the white qubits should be measured in basis $X$. In the right hand, By the removing rule the empty points are filled by blue qubits that should be measured in the $Z$-basis. } \label{ex2-c}
\end{figure}
In Figure (\ref{ex2-c}), we show the graph state corresponding to the color code on a hexagonal lattice. The degree of vertices in this graph is six and we can not match it to a square lattice. Therefore, as it is shown in Figure (\ref{ex2-c}), we use the uniformizing rule to any vertices of the graph. We denote the added qubits by white circles where they should be measured in the $X$-basis. After that, the graph can be matched on a square lattice, but there is still some empty points that should be filled by blue qubits measured in $Z$-basis. In this way we have a cluster state where all white qubits are measured in the $X$-basis and the blue qubits are measured in the $Z$-basis. The final state on the unmeasured qubits is the initial CC state.
\section{Appendix C}\label{C1}
It has been shown that twists of the toric code play role of non-abelian Ising anyons. In \cite{9}, author derived different properties of Ising anyons such as fusion rule and braiding rule for the twists. In this appendix, we briefly explain the fusion rule of Ising anyons for our model. To this end, we suppose a topological charge which is called $\sigma$ for each twist. Then, we show that this topological charge satisfies fusion properties of an Ising anyon. In an anyonic model with labels $\{a,b,c,...\}$, topological charge of an anyon, like "a", will be revealed in different phase factors of the wave function after braiding other anyons with the anyon $a$. Consequently, we should consider the effect of braiding different anyons of the TC model with each twist in our model. Since, It is known that if we braid two different anyons for twice, it will be equal to wind one of them around another one, we consider winding an anyon around twist in our model. \\

In the toric code model, winding an anyon around another one is equal to applying an operator corresponding to a closed string. However, since the topological charge of an anyon $e$ or $m$ changes in effect of transporting around the twist, there is no closed string of $m$ or $e$ anyons around a twist. Therefore, they can not play any role in determining the topological charge of the twist. However, there is also a fermion $\epsilon$ in the toric code model where $\epsilon=e\times m=m\times e$. By this fact,  since anyons $e$ and $m$ are converted to each other after transporting around the twist, it is possible to have a closed string of $\epsilon$ around a twist and we denote such an operation by $L_{\epsilon}$. Consequently, the topological charge of a twist can be determined by winding a fermion $\epsilon$. \\

Therefore, we create a pair of fermions $\epsilon$ and move one of them around a twist and then fuse with the first one. Since by operating $L_{\epsilon}$, a charge particle is converted to a flux particle, $L_{\epsilon}$ is equivalent with exchange of the flux and charge anyon which corresponds to the braiding operator $R_{e,m}$. It is clear that by repeating the operation $L_{\epsilon}$ we will exchange again the flux and charge anyons and it is also equal to the braiding operator $R_{m,e}$. By the fact that $R_{e,m} R_{m,e}=-1$, we conclude that $L_{\epsilon}^2 =-1$. It means that eigenvalues of the operator $L_{\epsilon}$ are $i$ or $-i$. such values shows different topological charge of the twists.\\

String operators in the new model are in three different kinds. We denote an open string of charge, flux and $\epsilon$ particles by $S_e$, $S_m$ and $S_{\epsilon}$, respectively. Furthermore, we denote a closed string of $e$ and $m$ particles by $L_e$ and $L_m$ where there is no such a closed string around a twist. We also have another closed string $L_{\epsilon}$ corresponding to a closed string of $\epsilon$ around a twist. In the following, we use these string operators to study fusion rules for twists:\\

\begin{figure}[t]
\center
\includegraphics[width=10cm,height=5cm,angle=0]{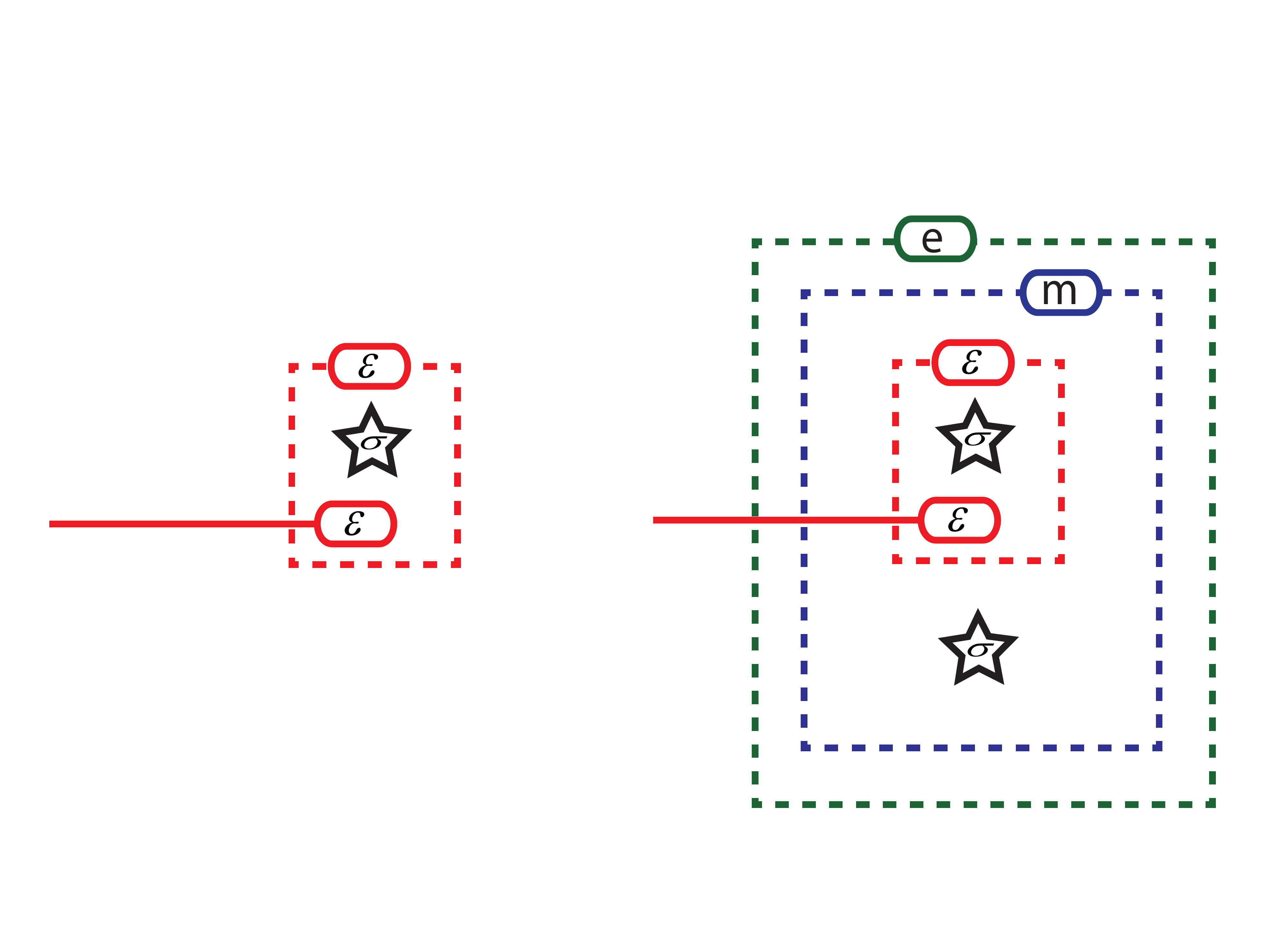}
\caption{(Color online) Each twist has been denoted by a star. Left: a closed string of a fermion $\epsilon$ can not characterize entering a fermion into the region involved by the red loop. Right: green and blue closed strings corresponding to winding charge and flux anyons around a pair of twists can characterize entering a fermion into the region involved by the blue and green loops. A red closed string corresponding to winding fermions $\epsilon$ around one of the twists can not characterize entering a fermion into the region involved by the red loop.} \label{fusion}
\end{figure}

Consider a region of the lattice which involves a twist $\sigma$. We can confirm existence of a twist in that region by measuring the operator $L_{\epsilon}$ where values $i$ or $-i$ will be derived, we have shown this operation by a red loop in the Figure (\ref{fusion}, left). Now, if we enter another fermion $\epsilon$ into the same region by applying a string operator $S_{\epsilon}$. Since $S_{\epsilon}$ commutes with $L_{\epsilon}$,  it is clear that the eigenvalue of $L_{\epsilon}$ does not change. it means that, total topological charge of the above region does not change. In other words, a fermion $\epsilon$ and a twist $\sigma$ have a total topological charge which is equal to $\sigma$. This result is equal to the fusion rule $\epsilon \times \sigma =\sigma $ which is one of the fusion rules of non-abelian Ising anyonic model where $\sigma $ is called Ising anyon.\\

It is also possible to explain another fusion rule of Ising anyons as $\sigma \times \sigma =1+\epsilon$. To this end, consider again the Figure (\ref{twist}), it is clear that if charge or flux anyons wind around a pair of twists $G$ and $\hat{G}$, they can create closed strings around those twists because their topological charges do not change, see the green loop in Figure (\ref{twist}). Consequently, it is possible to determine the total topological charge of a pair of twists by measuring the corresponding operators $L_e$ and $L_m$. By this fact, consider a region of the lattice containing a pair of twists and suppose that we measure the total charge of that region by winding a charge anyon and a flux anyon which are denoted by blue and green loops in Figure (\ref{fusion}, right), respectively. It is clear that the result of such operations will be equal to eigenvalue $+1$. it means that such loops do not measure any particles in this region and the total topological charge of the region is equal to vacuum and is also equal to the fusion rule $\sigma \times \sigma=1$. \\

Now, let us enter a fermion $\epsilon$ into the same region. Since the corresponding operator $S_{\epsilon}$ does not commutes with the loop operators $L_e$ and $L_m$, the result of measurement of these two loop operators will be equal to $-1$. It means that  two blue and green loops characterize presence of the $\epsilon$ fermion in the above region. Therefore, the total topological charge of the region is equal to a fermion $\epsilon$. Furthermore, we already proved $\sigma \times \epsilon =\sigma$, it means that existence of a fermion in the region does not change the topological charge of twists individually. We conclude that on the one hand, the topological charge of each twist is equal to $\sigma$ even in presence a fermion $\epsilon$ and on the other hand, the total topological charge of both twists is equal to $\epsilon$. Consequently, it means another fusion channel that $\sigma \times \sigma=\epsilon$. Finally there can be also both fusion channels as a degeneracy where $\sigma \times \sigma =1+\epsilon$.\\

\end{document}